\documentclass[prd,aps,floats,floatfix,nofootinbib,preprintnumbers]{revtex4-1}
\usepackage{amsmath}
\usepackage{amsfonts}
\usepackage{graphicx}
\usepackage{slashed}
\usepackage{dcolumn}
\usepackage{bm}
\usepackage{amssymb}
\usepackage{latexsym}
\usepackage{color}

\newcommand{\pbrac}[1]{\left( #1 \right)}
\newcommand{\tbrac}[1]{\left[ #1 \right]}
\newcommand{\cbrac}[1]{\left\{ #1 \right\}}

\begin{document}

\title{LHC Constraints on a Higgs Partner from an Extended Color Sector}

\author{R. Sekhar Chivukula}
\email[]{sekhar@msu.edu}
\affiliation{Department of Physics and Astronomy, Michigan State University, East Lansing, Michigan 48824, USA}
\preprint{MSUHEP-130425}
\author{Arsham Farzinnia}
\email[]{farzinnia@ibs.re.kr}
\affiliation{Center for Theoretical Physics of the Universe
, Institute for Basic Science (IBS), Daejeon 305-811, Republic of Korea}
\preprint{CTPU-14-02}
\author{Jing Ren}
\email[]{jingren2004@gmail.com}
\affiliation{Institute of Modern Physics and Center for High Energy Physics, Tsinghua University, Beijing 100084, China}
\author{Elizabeth H. Simmons}
\email[]{esimmons@msu.edu}
\affiliation{Department of Physics and Astronomy, Michigan State University, East Lansing, Michigan 48824, USA}

\date{\today}

\begin{abstract}
{We discuss the properties and LHC phenomenology of a potentially discoverable heavy scalar boson ($s$) that arises in the context of the renormalizable coloron model; the model also contains a light scalar, $h$, identifiable with the 125~GeV state discovered by the LHC.  These two scalar mass eigenstates are admixtures of a weak doublet gauge eigenstate  and a weak singlet gauge eigenstate.  A previous study set exclusion limits on the heavy $s$~scalar, using the stability of the scalar potential, unitarity, electroweak precision tests, LHC searches for the 125~GeV Higgs; it also briefly discussed the $\sqrt{s} = 7,8$~TeV LHC searches for a heavy Higgs.  In this work, we  show how the projected LHC sensitivity at $\sqrt{s} = 14$~TeV to the presence of a heavy Higgs and to the detailed properties of the 125~GeV Higgs will further constrain the properties of the new heavy $s$~scalar.   Since the renormalizable coloron model may contain spectator fermions to remove anomalies, we examine several representative scenarios with different numbers of spectator fermions.  Our results are summarized in plots that overlay the current exclusion limits on the $s$~boson with the projected sensitivity of the $\sqrt{s} = 14$~TeV LHC to the new state.  We find that the upcoming LHC searches should be sensitive to an $s$~scalar of mass less than 1~TeV for essentially all of the model parameter space in which the $h$~state differs from the Higgs boson of the SM.  More precisely, unless the mixing between the weak doublet and weak singlet gauge-eigenstate scalars is zero, the 14~TeV LHC will be sensitive to the presence of the non-standard heavy $s$~state that is characteristic of the renormalizable coloron model.}
 \end{abstract}

\maketitle

\section{Introduction}\label{intro}

An extended color sector is an integral part of many theories with new dynamics that attempt to address unresolved challenges with the ordinary standard model (SM). Models including topcolor \cite{Hill:1991at}, the flavor-universal coloron \cite{Chivukula:1996yr}, chiral color \cite{Frampton:1987dn}, chiral color with unequal gauge couplings \cite{Martynov:2009en} and flavor non-universal chiral color \cite{Frampton:2009rk} belong to this class of theories beyond the SM.

In such models, the minimal extension of the color sector consists of enhancing the color gauge group to $SU(3)_{1c} \times SU(3)_{2c}$, which is spontaneously broken to the diagonal $SU(3)_{c}$ subgroup. Ordinary QCD is then identified with this diagonal unbroken gauge group, containing the usual massless gluon color-octet. The spontaneous symmetry breaking produces, furthermore, a set of massive color-octet gauge bosons, which we generically refer to as colorons. Spontaneous breaking of the enhanced color symmetry may be facilitated in a renormalizable manner---in conjunction with electroweak symmetry breaking---by means of an enlarged scalar sector \cite{Hill:1993hs,Dicus:1994sw,Chivukula:1996yr,Bai:2010dj,Chivukula:2013xka}. As a consequence, new (colored and uncolored) scalar degrees of freedom are present in this formalism; most notably there is a color-singlet (heavy) Higgs-like scalar in addition to the electroweak Higgs boson. Moreover, cancelation of potential anomalies---brought forth by the possible chiral couplings of the ordinary quarks to colorons---may require the existence of additional spectator fermions. Hence, if this formalism corresponds to the correct description of nature, a rich spectrum of new (colored) scalar, fermionic, and vector particles with novel properties are predicted, the discovery of which may lie within the reach of the LHC.

A first complete study of hadron collider production of colorons at next-to-leading order was presented in \cite{Chivukula:2011ng,Chivukula:2013xla}. The scalar sector of the renormalizable coloron model has also been constrained in a previous analysis \cite{Chivukula:2013xka}, by imposing limits from the stability of the potential, unitarity, electroweak precision tests, and properties of the 125~GeV Higgs-like scalar, discovered at the LHC \cite{Aad:2012tfa,Chatrchyan:2012ufa}. Furthermore, the latter study presented a brief discussion of the LHC heavy Higgs searches at $\sqrt s = 7,8$~TeV \cite{CMSHeavyH,ATLASHeavyH}, applied to the additional CP-even scalar of the renormalizable coloron model.

In the present paper, we further explore the properties of the additional scalar boson, with a mass up to 1~TeV, arising from the minimally extended color sector. Specifically, we discuss the implications of the expected reach of the $\sqrt s = 14$~TeV LHC with an integrated luminosity of 300 and 3000 fb$^{-1}$ for the heavy Higgs \cite{ATLASHeavyS,Brownson:2013lka,CMSHeavyS}, and at 300 fb$^{-1}$ for the detailed properties of the discovered 125~GeV Higgs \cite{ATLASHproj}. If no signal of a heavy scalar is found nor any deviations from the current measurements of the 125~GeV Higgs detected by the $\sqrt s = 14$~TeV LHC with the quoted luminosities, these projections will impose complimentary constraints at 95\%~C.L. on the model's parameter space. We present the plots depicting the current exclusion limits, as well as the projected 14~TeV exclusion limits, for three selected scenarios with 0, 1, and 3 spectator fermion generations, varying the remaining free parameters within each model to demonstrate the sensitivity of the analyses to these variables.  We show thatthe large regions of the parameter space are sensitive to future probing by the LHC.

In Section~\ref{review}, we briefly describe the formalism of the renormalizable coloron model  \cite{Hill:1993hs,Dicus:1994sw,Chivukula:1996yr,Bai:2010dj,Chivukula:2013xka}, review the particle content, and introduce relevant notation. Section~\ref{sHiggs} is devoted to setting up an effective Lagrangian utilized for the subsequent phenomenological study of the heavy $s$ scalar, arising from this minimally extended color sector. The phenomenology of the LHC heavy scalar searches and projections, along with other considerations such as the total width of the heavy $s$ boson in the renormalizable coloron model, are presented in Section~\ref{sPheno} and summarized in plots, covering the free parameter space for representative scenarios. We further discuss the results in Section~\ref{disc} for the different models and their ranges of free parameters, and we demonstrate that all scenarios may be fully explored by the LHC. If no heavy scalar is discovered at or below a 1~TeV mass, the searches will exclude the renormalizable coloron model within the TeV range. Furthermore, we elaborate on distinguishing a potentially discovered heavy scalar, originating from the renormalizable coloron model, from other color-singlet ``Higgs portal'' scenarios \cite{Schabinger:2005ei,Barbieri:2005ri,Patt:2006fw}. Finally, we present our conclusions in Section~\ref{concl}.

\section{Review of the Renormalizable Coloron Model}\label{review}

We start by briefly reviewing the formalism of the minimal renormalizable coloron model \cite{Hill:1993hs,Dicus:1994sw,Chivukula:1996yr,Bai:2010dj,Chivukula:2013xka}, and summarizing the gauge, scalar, and fermionic sectors of the theory.\footnote{Throughout this paper, we closely follow the notation introduced in \cite{Chivukula:2013xka}.} In the renormalizable coloron model, the color gauge group of the standard model (SM) is extended to $SU(3)_{1c} \times SU(3)_{2c}$, which is spontaneously broken to the diagonal $SU(3)_{c}$ subgroup, and the latter is identified with the ordinary QCD. Schematically, we have
\begin{equation}\label{gaugegroup}
SU(3)_{1c} \times SU(3)_{2c} \times SU(2)_{L} \times U(1)_{Y} \; \longrightarrow \; SU(3)_{c} \times U(1)_{\text{EM}}.
\end{equation}

Spontaneous symmetry breaking in this enlarged color sector, along with spontaneous breaking of the SM electroweak gauge group, proceeds via an extended scalar sector. In addition to the usual SM Higgs doublet, $\phi$, the scalar sector contains a $(3,\bar{3})$ scalar, $\Phi$, that is bi-fundamental under the $SU(3)_{1c} \times SU(3)_{2c}$ color group, but singlet under the electroweak interactions. Component-wise these two scalars may be written as \cite{Bai:2010dj,Chivukula:2013xka}
\begin{align}
\phi=&\, \frac{1}{\sqrt{2}}
\begin{pmatrix} i\sqrt{2}\,\pi^+ \\ v_h+h_0+i\pi^0 \end{pmatrix} \ , \label{phi} \\
\Phi =&\, \frac{1}{\sqrt{6}} \pbrac{v_{s} + s_{0} + i {\cal A}} {\cal I}_{3\times 3} + \pbrac{G^a_H + i G^a_G}t^a \qquad \pbrac{t^a \equiv \lambda^a/2}\label{Phi} \ ,
\end{align}
where $h_0$ is the SM Higgs boson with the corresponding vacuum expectation value (VEV) $v_h=246$~GeV, $\pi^{0,\pm}$ are the electroweak Nambu-Goldstone bosons, and $\lambda^{a}$ are the Gell-Mann matrices. As explained in \cite{Chivukula:2013xka}, the fields $s_{0}$ and $\cal A$ are, respectively, CP-even and CP-odd gauge-singlet scalars, whereas $G_{H}^{a}$ denotes a set of scalars that transform as a color-octet under the $SU(3)_{c}$ color group. In addition, the $\Phi$ field contains a set of massless colored Nambu-Goldstone bosons, $G_G^a$. The CP-even gauge-singlet, $s_{0}$, develops a non-zero VEV, $v_{s}$, resulting in spontaneous breaking of the extended color symmetry.

It has been shown in \cite{Chivukula:2013xka} that the most general renormalizable scalar potential is of the form
\begin{equation}\label{pot}
\begin{split}
V(\phi,\Phi) = &\, \frac{\lambda_s}{6}\pbrac{{\rm Tr}\tbrac{\Phi^\dagger \Phi}}^2 + \frac{\kappa_s}{2} {\rm Tr}\tbrac{\pbrac{\Phi^\dagger \Phi}^{2}} -\frac{\lambda_s + \kappa_s}{\sqrt{6}}\,r_\Delta v_{s} \pbrac{{\rm det}\Phi + {\rm h.c.}}  -\frac{\lambda_s + \kappa_s}{6} \, v_s^{2} \pbrac{1 - r_{\Delta}} {\rm Tr}\tbrac{\Phi^\dagger \Phi} \\
& +\frac{\lambda_h}{6}\pbrac{\phi^\dagger \phi - \frac{v^2_h}{2}}^2 + \lambda_m\pbrac{\phi^\dagger \phi - \frac{v^2_h}{2}} \pbrac{{\rm Tr}\tbrac{\Phi^\dagger \Phi} - \frac{v^2_s}{2}} \ ,
\end{split}
\end{equation}
with $\lambda_{h}$, $\lambda_{m}$, $\lambda_{s}$, $\kappa_{s}$, and $r_{\Delta}$ all dimensionless couplings. Moreover, it was demonstrated that this potential is bounded from below, with a global minimum coinciding with the VEVs shown above, given the conditions
\begin{equation}\label{stab}
\lambda_h > 0 \ , \qquad \lambda_s^{\prime} > 0 \ , \qquad \kappa_{s} > 0 \ , \qquad \lambda_m^2 < \frac{1}{9} \lambda_h \lambda_s^{\prime} \ , \qquad 0 \le r_{\Delta} \le \frac{3}{2} \ ,
\end{equation}
with $\lambda_{s}^{\prime} \equiv \lambda_{s} + \kappa_{s}$.

In the mass eigenstate basis, the CP-even scalars $h_0$ and $s_0$ in \eqref{phi} and \eqref{Phi} are mixed due to the $\lambda_m$ term in the potential \eqref{pot} and due to the fact that they both develop non-zero VEVs. All the other scalar masses remain diagonal in both mass and field eigenstate bases. The corresponding mass eigenstates $h$ and $s$ composed of the original CP-even scalars may be defined using an orthogonal rotation
\begin{equation}\label{massbasis}
\begin{pmatrix} h_0\\ s_0 \end{pmatrix}
=  \begin{pmatrix} \cos\chi & \sin\chi \\ -\sin\chi & \cos\chi \end{pmatrix} \begin{pmatrix} h \\ s \end{pmatrix} \ ,
\end{equation}
with the mixing angle, $\chi$, given by
\begin{equation}\label{chi}
\cot 2\chi \equiv \frac{1}{6\lambda_m} \tbrac{ \lambda_s^{\prime}\pbrac{1-\frac{r_{\Delta}}{2}} \frac{v_s}{v_h} - \lambda_h \frac{v_h}{v_s} } \ .
\end{equation}
The diagonal scalar masses then read
\begin{align}
m_{h,s}^{2} =& \frac{1}{6} \cbrac{ \lambda_h v_h^2 + \lambda_s^{\prime} v_s^2\pbrac{1-\frac{r_{\Delta}}{2}} \pm \tbrac{\lambda_h v_h^2 - \lambda_s^{\prime} v_s^2\pbrac{1-\frac{r_{\Delta}}{2}}}\sec 2\chi } \ , \label{mhs} \\
m_{\cal A}^{2} =&\, \frac{v_{s}^{2} }{2} \, r_{\Delta} \lambda_{s}^{\prime} \ , \qquad m_{G_{H}}^{2} = \frac{1}{3} \tbrac{v_{s}^{2} \, \kappa_{s}+ 2 m_{\cal A}^{2}} \ . \label{mAGH}
\end{align}
The $h$ scalar is assumed to be the lighter of these two CP-even degrees of freedom, and is identified with the discovered 125 GeV Higgs-like state at the LHC \cite{Aad:2012tfa,Chatrchyan:2012ufa}. For small mixing angle values---motivated by the experimental constraints \cite{Chivukula:2013xka}---the $h$ is more ``SM-like", whereas the heavier $s$ is more ``singlet-like" (c.f. \eqref{massbasis}). They are both capable of interacting with the ordinary SM particles, albeit with suppressed tree-level couplings proportional to $\cos \chi$ and $\sin \chi$, respectively, as compared with a pure SM Higgs. Analyzing the phenomenology of the heavier scalar, $s$, constitutes the main subject of the present study.

Examining \eqref{mhs} and \eqref{mAGH} together with the conditions \eqref{stab}, one further notes the following mass relations  \cite{Chivukula:2013xka}
\begin{equation}\label{masscond}
m_s \ge \frac{1}{3}\, m_{\mathcal A} \quad \pbrac{\text{for}\; \sin \chi \to 0} \ , \qquad m_{G_{H}} \geq \sqrt \frac{2}{3} \, m_{\mathcal A} \ .
\end{equation}
A scalar color-octet mass, $m_{G_{H}}$, ranging from  50 to 125~GeV is already excluded by the Tevatron searches \cite{Aaltonen:2013hya}; hence, following \cite{Chivukula:2013xka}, we treat this color-octet set as being heavier than the discovered $h$~scalar ($m_{G_{H}} > m_{h}=125$~GeV) throughout our analyses.

The spontaneous breaking of the enhanced color symmetry to its diagonal subgroup produces, in addition to the usual massless $SU(3)_{c}$ gluons, a set of massive color-octet vector bosons, generically referred to as colorons. Colorons obtain their mass by ``eating" the colored Nambu-Goldstone bosons, $G_{G}^{a}$ in \eqref{Phi}; the mass is given by the expression \cite{Bai:2010dj,Chivukula:2013xka}
\begin{equation}\label{MC}
M_C = \sqrt{\frac{2}{3}}\frac{g_s\, v_{s}}{\sin 2\theta_c} \qquad
\pbrac{\sin\theta_c \equiv \frac{g_{s_1}}{\sqrt{g_{s_1}^2+g_{s_2}^2}}}\ .
\end{equation}
Here, $\theta_c$ represents the mixing angle in the orthogonal matrix rotating the $SU(3)_{1c}$ and $SU(3)_{2c}$ gauge eigenstates (with the corresponding couplings $g_{s_1}$ and $g_{s_2}$) into the gluon and coloron mass eigenstates. Furthermore, $g_{s}$ denotes the QCD gauge coupling, which may be expressed as the combination \cite{Chivukula:2011ng,Chivukula:2013xla,Chivukula:2013xka}
\begin{equation}\label{gs}
\frac{1}{g_s^2} = \frac{1}{g_{s_1}^2}+\frac{1}{g_{s_2}^2}\ .
\end{equation}
Note that Tevatron and current LHC searches constrain the coloron mass, $M_{C}$, to be at least in the TeV region \cite{Simmons:1996fz,Bertram:1998wf,ATLAS:2012pu,ATLAS:2012qjz,Chatrchyan:2013qha,CMS:kxa}. 

As described in \cite{Chivukula:2013xka}, in the context of the renormalizable coloron model, the fermionic matter sector needs modifications as well, possibly including the addition of heavy spectator quarks \cite{Frampton:1987dn,Frampton:1987ut,Cvetic:2012kv,Chivukula:2013xla}. These extra fermionic degrees of freedom can be necessary to cancel potential anomalies of the theory (arising once the couplings of the quarks to the $SU(3)_{1c} \times SU(3)_{2c}$ color gauge group are chosen to be chiral \cite{Chivukula:2011ng, Chivukula:2013xla}), and must have the opposite chirality charges as the ordinary quarks.

Allowing for the spectator fermions to obtain their mass via Yukawa interactions with the $\Phi$ scalar, one deduces
\begin{equation}\label{MQ}
M_{Q} = \frac{y_{Q}}{\sqrt 6}\, v_{s} \ ,
\end{equation}
where, for simplicity, a flavor-universal Yukawa coupling, $y_Q$---and hence spectator fermion mass scale---is assumed. Furthermore, it is conjectured \cite{Chivukula:2013xka} that the spectator flavors have the same electric charges as their corresponding quark partners, and that they are vectorial under the electroweak interactions.\footnote{Spectator quarks that are chiral under the electroweak interactions necessitate the existence of additional ``lepton-like" spectators to cancel further introduced anomalies.} Flavor-changing couplings of the colorons result in strong constraints on potential mixing between these extra fermionic states and the ordinary quarks, rendering such a mixing negligible \cite{Chivukula:2013kw}.

In addition to those assumptions regarding the properties of the heavy spectator fermions and their interactions, the chiral nature of a specific model determines the number of necessary spectator flavors to cancel potential anomalies \cite{Cvetic:2012kv,Chivukula:2013xka}. No spectators are required when all quarks are vectorially charged under the extended color group, since no anomaly will be introduced. If, however, the chiral couplings of the third quark generation are chosen to be opposite to those of the first two generations, then one spectator generation (one up-like and one down-like spectator) is necessary. Flavor-universal chiral interactions of the quarks under the enlarged color gauge group require, on the other hand, three generations of spectator fermions to cancel all anomalies. Following \cite{Chivukula:2013xka}, we shall study the phenomenological results for three cases described above, where 0, 1, or 3 spectator quark generations may be present in the theory. 

In summary, the renormalizable coloron model extends the SM color gauge group minimally (c.f. \eqref{gaugegroup}), while appropriately enlarging the scalar sector (to also accommodate spontaneous breaking of the enhanced color symmetry), as well as the fermionic matter sector (to cancel potentially introduced anomalies). Setting the electroweak VEV at $v_h = 246$~GeV and the mass of the $h$ Higgs at $m_{h}=125$~GeV, it adds eight new free parameters to the usual SM,\footnote{This is, of course, true once the aforementioned properties and interactions of the spectator fermions are assumed.} which may be taken as the set \cite{Chivukula:2013xka}
\begin{equation}\label{freepar}
\cbrac{v_{s}, \sin \chi, m_{s}, m_{\cal A}, m_{G_{H}}, M_{C}, M_{Q}, N_Q} \ ,
\end{equation}
with $N_Q$ the number of spectator fermion generations, which we keep as a free parameter throughout our analyses.

For completeness, we exhibit the explicit dependence of the Lagrangian parameters on the set \eqref{freepar}, which can be straightforwardly derived from \eqref{mhs}, \eqref{mAGH}, \eqref{MC}, and \eqref{MQ},
\begin{equation}\label{conversions}
\begin{split}
\lambda_{h} = &\, \frac{3}{2}\frac{m_{h}^{2}+m_{s}^{2}+\pbrac{m_{h}^{2}-m_{s}^{2}}\cos 2\chi}{v_{h}^{2}} \ , \qquad \lambda_{m} = -\frac{1}{2}\frac{m_{h}^{2}-m_{s}^{2}}{v_{h} v_{s}}\sin 2\chi \ , \\
\lambda_{s}^{\prime} = &\, \frac{1}{2}\frac{2m_{\mathcal A}^{2}+3\pbrac{m_{h}^{2}+m_{s}^{2}}-3\pbrac{m_{h}^{2}-m_{s}^{2}}\cos 2\chi}{v_{s}^{2}} \ , \qquad \kappa_{s} = \frac{3m_{G_{H}}^{2}-2m_{\mathcal A}^{2}}{v_{s}^{2}} \ , \\
r_{\Delta} = &\, \frac{4 m_{\mathcal A}^{2}}{2m_{\mathcal A}^{2}+3\pbrac{m_{h}^{2}+m_{s}^{2}}-3\pbrac{m_{h}^{2}-m_{s}^{2}}\cos 2\chi} \ , \qquad \sin 2\theta_{c} = \sqrt \frac{2}{3} \frac{g_{s} \, v_{s}}{M_{C}} \ , \qquad y_{Q} =\sqrt 6 \frac{M_{Q}}{v_{s}} \ .
\end{split}
\end{equation}
It is worth noting that the sign of $\lambda_m$ (i.e. an attractive or repulsive interaction between the scalar fields $\phi$ and $\Phi$ in \eqref{pot}) does not affect the stability of the potential (c.f. \eqref{stab}), and is reflected only in the sign of the parameter $\sin \chi$ in \eqref{conversions}. As we shall demonstrate, however, this sign has notable effects on the phenomenology of the heavy Higgs-like scalar,~$s$.

\section{The Heavy $\textit{\normalsize{s}}$ Scalar}\label{sHiggs}

As explained in the previous section, the heavy $s$~scalar is an admixture of the SM $h_0$ Higgs and the gauge-singlet $s_0$ boson (c.f. \eqref{massbasis}), and is mostly singlet-like for the small mixings favored by the LHC 125~GeV signal data and the electroweak precision tests \cite{Chivukula:2013xka}. This heavy scalar is, nevertheless, capable of interacting weakly with the SM particles, while its tree-level coupling is suppressed by the mixing angle factor, $\sin \chi$, with respect to that of the usual SM Higgs boson, $h_0$. The LHC heavy Higgs searches \cite{CMSHeavyH,ATLASHeavyH} and projected 14~TeV exclusion limits \cite{ATLASHeavyS,Brownson:2013lka,CMSHeavyS} in the vector boson channels can, thus, be utilized to investigate and constrain the properties of the $s$ boson arising from the extended color sector.

To this end, we set up an effective Lagrangian \cite{Chivukula:2013xka} to parametrize all the relevant decay channels of the heavy $s$~Higgs at the LHC
\begin{equation}
\label{heavyLeff}
\begin{split}
\mathcal{L}^s_{\text{eff}} =&\,
   c^s_V \frac{2m_W^2}{v_h}\,s\,W_{\mu}^{+}W^{-\mu}
  +c^s_V \frac{m_Z^2}{v_h}\,s\,Z_{\mu}Z^{\mu}
  -c^s_t \frac{m_t}{v_h}\,s\,\bar{t}t
  -c^s_b \frac{m_b}{v_h}\,s\,\bar{b}b
  -c^s_\tau \frac{m_\tau}{v_h}\,s\,\bar{\tau}\tau
  -c^s_c \frac{m_c}{v_h}\,s\,\bar{c}c\\
&
  +c^s_g\frac{\alpha_s}{12\pi{v_h}}\,s\,G^a_{\mu\nu}G^{a\mu\nu}
  +c^s_h \,s\,h h
  +c^s_\mathcal{A} \,s\,\mathcal{A}\mathcal{A}
  +c^s_{G_H} \,s\,G_H^a G_H^a \ .
\end{split}
\end{equation}
In addition to the tree-level SM gauge boson and heavy fermion channels, this effective Lagrangian describes the one-loop induced $s$~decay into a pair of gluons, as well as its tree-level decays into pairs of 125~GeV $h$~Higgses, $\mathcal{A}$ pseudoscalars, and $G_H^a$ scalar color-octets.\footnote{For spectator quarks with the same electric charge as the ordinary quarks, $s$~decay into di-photons is a small contribution to the total decay width and is not relevant to our analysis.} Given the TeV lower bounds on the coloron mass \cite{Simmons:1996fz,Bertram:1998wf,ATLAS:2012pu,ATLAS:2012qjz,Chatrchyan:2013qha,CMS:kxa}, we have neglected a potential decay of the $s$~boson into a pair of colorons in \eqref{heavyLeff}. Furthermore, we focus our attention on the case where the spectator fermions are heavy, with $M_Q$ in the TeV range \cite{CMS:2013tda,ATLAS:2013ima,Ellis:2014dza}, such that the $s$~decay into a pair of spectator fermions is also kinematically prohibited. In this regime, the $s$~boson production and decay amplitudes depend crucially on the number of spectator quarks, but are relatively insensitive to the precise mass of these fermions. Since the coupling of the $s$~boson to a coloron or spectator fermions is proportional to the coloron's or spectator's mass (\eqref{MC} and \eqref{MQ}, respectively), the contributions of the spectator fermions and colorons to $s$~boson production via gluon fusion do not ``decouple" in the heavy mass limit. The production cross section is, therefore, not significantly suppressed in this regime.

In case of the $s$~decay into a SM final state, such as $gg$ or $\bar \tau \tau$, the dimensionless coefficient $c^s_i$ in \eqref{heavyLeff} represents the appropriate deviation of the corresponding coupling from its SM value (i.e., if $s$ were replaced by the usual SM Higgs boson, $h_{0}$). The tree-level coefficients are, in this case, given by
\begin{equation}\label{csSM}
c^s_V=c^s_t=c^s_b=c^s_\tau=c^s_c=\sin\chi \ ,
\end{equation}
whereas, the coefficient associated with the one-loop effective gluon coupling (with colorons, spectators, and scalar color-octets running in the loop, in addition to the ordinary SM heavy quarks) may be parametrized by \cite{Chivukula:2013xka}
\begin{align}
c^s_g=&\, \sin\chi \,\hat{c}_g^{s, \text{SM}}+\cos\chi \, \delta c^s_g \ ,\label{eq:cgs}\\
\hat{c}_g^{s, \text{SM}}\equiv&\, A_F(\tau^s_t)+A_F(\tau^s_b)+A_F(\tau^s_c) \ ,\label{cgsSM}\\
\delta c^s_g\equiv&-3\frac{v_{h}}{v_s}\left[6A_V(\tau^s_C)+6\left(1+\frac{m_s^2-\frac{2}{3}m_\mathcal A^2}{2m_{G_H}^2}\right)A_S(\tau^s_{G_H})-\frac{2 N_Q}{3}A_F(\tau^s_Q)\right]\label{delcgs} \ .
\end{align}
Here, $\tau^s_i\equiv \dfrac{m_s^2}{4m_i^2}$, the subscript $C$ ($Q$) represents the coloron (spectator), $N_Q$ is the number of spectator fermion generations, and the vector, fermion, and scalar loop form factors are defined as
\begin{equation}\label{formfact}
\begin{split}
A_V(\tau)\equiv&\,\frac{1}{8\tau^2}\tbrac{3\tau+2\tau^2- 3(1-2\tau)f(\tau)} \ , \quad
A_F(\tau)\equiv \frac{3}{2\tau^2}\tbrac{\tau- (1-\tau)f(\tau)} \ , \quad
A_S(\tau)\equiv\frac{1}{8\tau^2}\tbrac{\tau-f(\tau)} \ , \\
f(\tau) \equiv&\,
\begin{cases}
\arcsin^{2}\sqrt{\tau} \qquad \qquad \qquad \qquad \; \,\tau\leq1\\
-\frac{1}{4}\tbrac{\log \frac{1+\sqrt{1-\tau^{-1}}}{1-\sqrt{1-\tau^{-1}}}-i\pi}^2 \qquad \tau>1
\end{cases} \ .
\end{split}
\end{equation}
It is important to notice the sign in \eqref{eq:cgs} associated with $\sin \chi$: an attractive or repulsive interaction between the scalars $\phi$ and $\Phi$ as parametrized by the sign of $\lambda_m$ in the potential \eqref{pot}---see also \eqref{conversions}.  This may result in a constructive or destructive interference between the two terms. In addition, a possible cancellation between the spectator and coloron contributions may occur in \eqref{delcgs}, depending on the chosen number of spectator generations, while the scalar term can be of either sign. This complicated interplay between the various parameters lead to non-trivial phenomenological consequences for the $s$ boson that will be discussed in Section~\ref{sPheno}. The remaining dimensionful scalar coupling coefficients were also derived in \cite{Chivukula:2013xka}, and we list them here for completeness
\begin{align}
c_h^s
=&\, -\frac{\sin\chi\cos\chi}{2v_hv_s}\left[v_h\left(\frac{m_{\mathcal{A}}^2}{3}+2m_h^2+m_s^2\right)\sin\chi+v_s(2m_h^2+m_s^2)\cos\chi\right] \ ,\label{chs}\\
c_\mathcal{A}^s
=&\,-\frac{m_\mathcal{A}^2+m_s^2}{2v_s}\cos\chi \ , \label{cAs}\\
c_{G_H}^s=&\, -\frac{m_s^2+2m_{G_H}^2-\frac{2}{3}m_\mathcal{A}^2}{2v_s}\cos\chi \label{cGHs} \ .
\end{align}

Armed with the effective Lagrangian \eqref{heavyLeff}, in the following section we shall adapt the LHC heavy Higgs searches and projections in the vector boson channels to study and constrain the properties of the $s$~boson.

\section{Phenomenology of the heavy $\textit{\normalsize{s}}$ Higgs}\label{sPheno}

The LHC heavy Higgs searches \cite{CMSHeavyH,ATLASHeavyH} and projected 14~TeV exclusion limits \cite{ATLASHeavyS,Brownson:2013lka,CMSHeavyS} generally assume that the narrow-width approximation is valid. Before applying these studies to examine the $s$~boson's properties, it is, therefore, necessary to confirm the validity of this approximation in the context of the heavy $s$~scalar originating from an extended color sector.

Using the constructed effective Lagrangian \eqref{heavyLeff} and the coefficients \eqref{csSM}-\eqref{cgsSM}, the total width of the $s$~boson as a function of its mass, $\Gamma_s^{\text{TOT}}(m_s)$, is given by \cite{Chivukula:2013xka}
\begin{equation}\label{Ctots}
\frac{\Gamma_s^{\text{TOT}}(m_s)}{\Gamma_{h_{0}}(m_s)} =\left|\frac{c_g^s}{\hat{c}_g^{s,\textrm{SM}}}\right|^2\textrm{BR}^{\textrm{SM}}_{gg}
+\sin^2\chi\tbrac{\textrm{BR}^{\textrm{SM}}_{VV}+\textrm{BR}^{\textrm{SM}}_{\bar{f}f}}+\frac{\Gamma(s\to hh)+\Gamma(s\to\mathcal{A}\mathcal{A})+8\, \Gamma(s\to G^a_HG^a_H)}{\Gamma_{h_{0}}(m_s)} \ ,
\end{equation}
where $\Gamma_{h_{0}}(m_s)$ denotes the total decay width of the SM $h_{0}$ Higgs when $m_{h_{0}} = m_s$ (c.f. \eqref{massbasis}). The expression \eqref{Ctots} takes into account the decay of $s$ into the usual SM final states, as well as into the additional scalar degrees of freedom present in the theory, such as the pseudoscalar $\mathcal{A}$ and the scalar color-octet $G_{H}^{a}$. Inserting the coefficients \eqref{chs}-\eqref{cGHs}, the corresponding widths of the $s$~decay into scalar pairs are defined as
\begin{equation}
\Gamma(s\to i i)=\frac{(c_i^s)^2}{8\pi \,m_s}\sqrt{1-\frac{4m_i^2}{m_s^2}} \qquad \pbrac{i=h, \mathcal{A}, G_H^a} \ .
\end{equation}

\begin{figure}
\includegraphics[width=.329\textwidth]{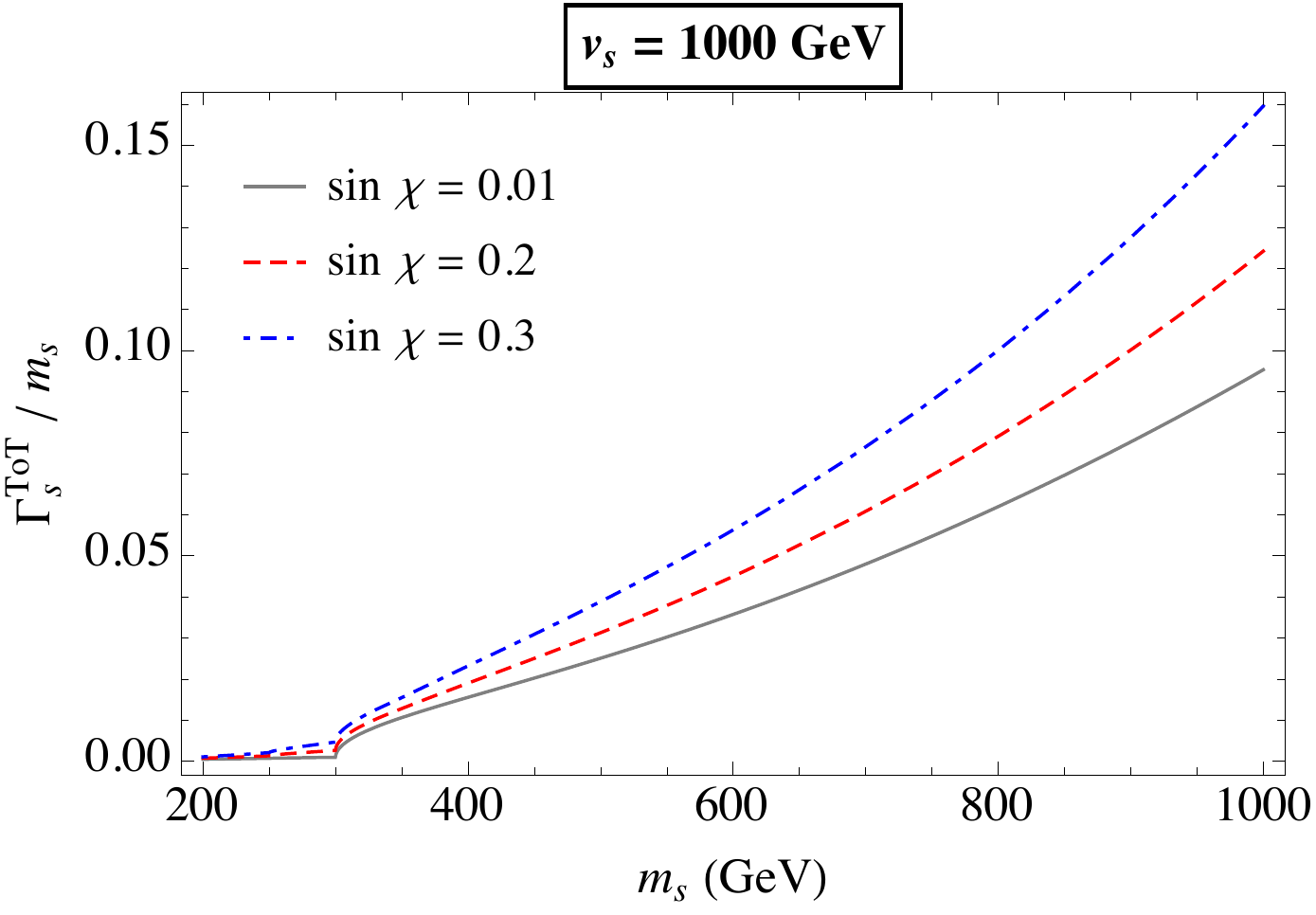}
\includegraphics[width=.329\textwidth]{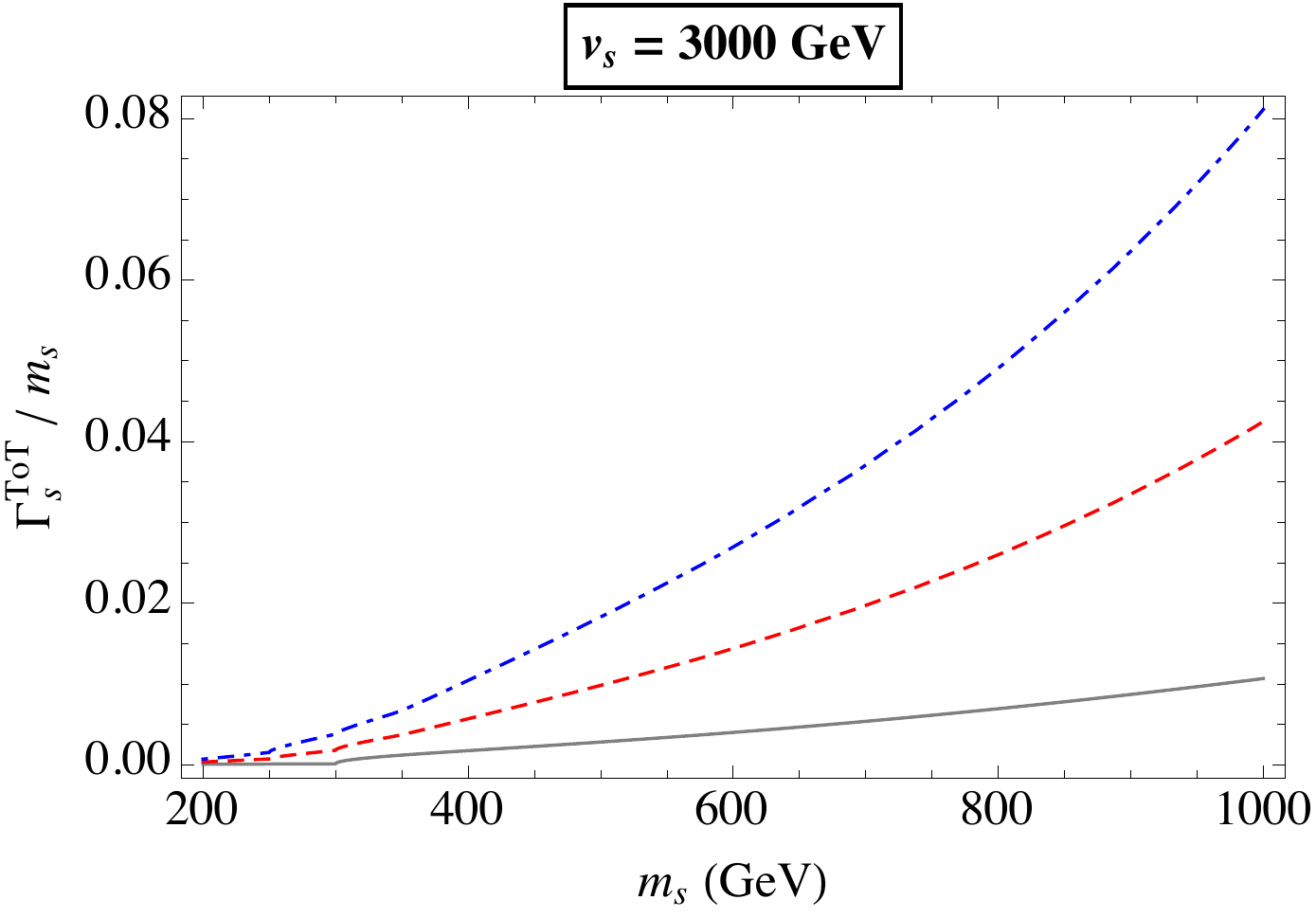}
\includegraphics[width=.329\textwidth]{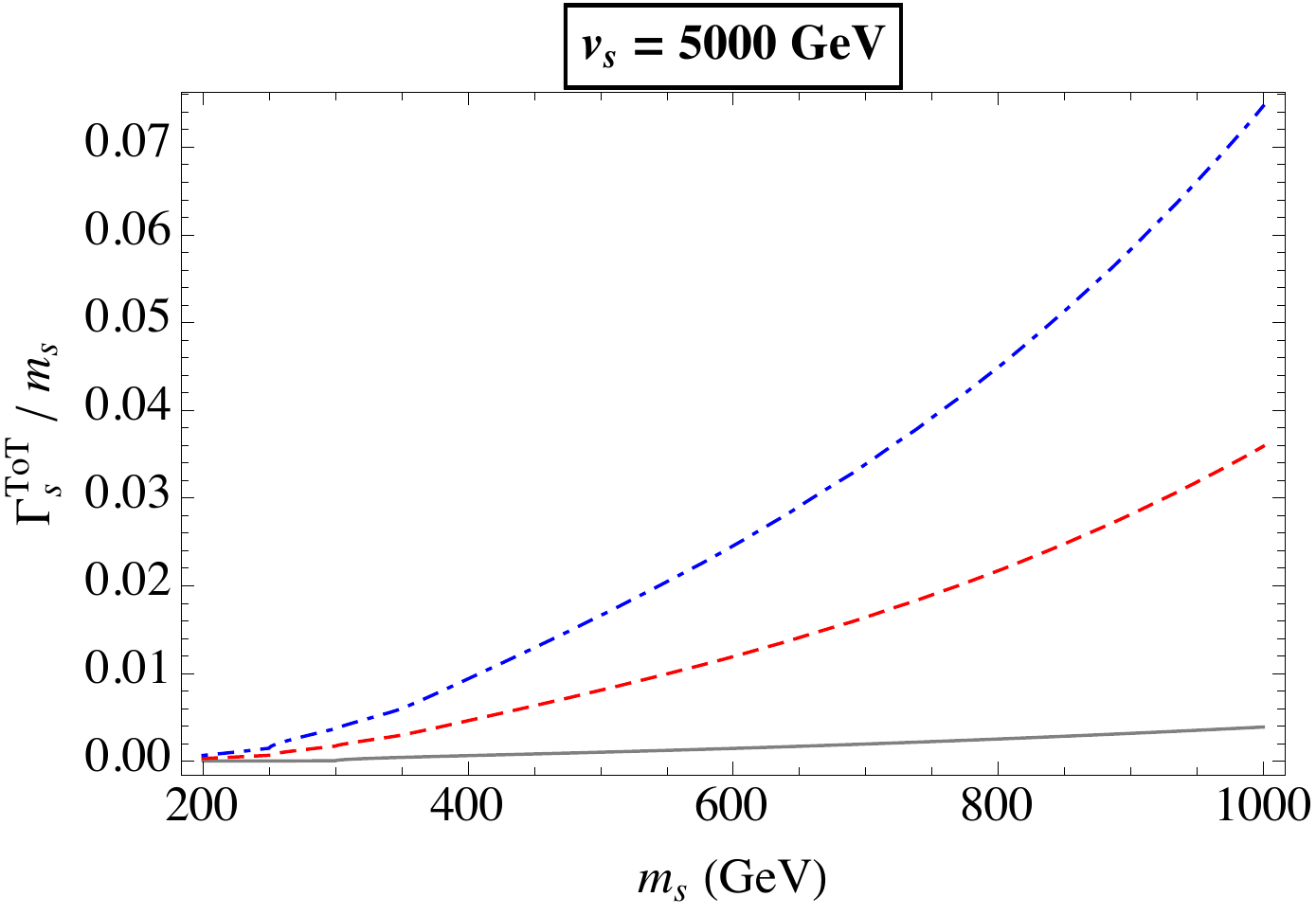}
\caption{The ratio of the $s$~boson's total width to its mass as a function of its mass for the range $200 \leq m_{s} \leq 1000$~GeV, exhibiting the validity of the narrow-width approximation. The panels correspond to three benchmark values of the singlet VEV, $v_{s}$, each displaying this ratio for three different mixing angles, $\sin \chi$, as motivated by the experimental searches \cite{Chivukula:2013xka}. A ``light'' pseudoscalar ($m_{\mathcal{A}}=50$~GeV) and scalar color-octet ($m_{G_{H}}=150$~GeV) have been chosen to make the width as broad as possible for the entire range of $m_{s}$, demonstrating the ``worst-case'' scenarios. The ratio is, furthermore, insensitive to the (TeV range) coloron and spectator masses, and the number of spectator fermion generations.}
\label{NWA}
\end{figure}

The narrow-width approximation concerns, specifically, the ratio of the total width of the heavy $s$~boson to its mass ($\Gamma_{s}^{\text{TOT}}/m_{s}$), which must remain small over the entire relevant range of the mass. This ratio has been plotted in Fig.~\ref{NWA} for three selected values of the singlet VEV, $v_{s}$, covering a heavy $s$~boson up to $m_{s}=1$~TeV. In each panel, three representative values of the mixing angle have been displayed, as motivated by the experimental bounds \cite{Chivukula:2013xka}. In all plots, the pseudoscalar and the scalar color-octet have been chosen to be light enough to contribute to the $s$~width, making the total width as broad as possible for the mass range of interest, with $m_{\mathcal{A}}=50$~GeV and $m_{G_{H}}=150$~GeV.\footnote{We have neglected an \textit{off-shell} decay of $s$ into pairs of $h$, $\mathcal A$ and $G_{H}^{a}$ in computing its total width. An off-shell decay of $s$ into the usual SM pairs (e.g. $\bar t t$) is, however, taken into account.} Hence, the plots depict a scenario in which decay to a pair of $\mathcal{A}$-bosons contributes to the $s$~width. The heavy TeV-range coloron and spectator masses render the ratio insensitive to the precise values of $M_{C}$ and $M_{Q}$ parameters. Sensitivity to the number of spectator fermion generations, $N_{Q}$, is also negligible. One can deduce from Fig.~\ref{NWA} that the $s$~boson's total width remains relatively narrow in the entire mass range of interest for all possible scenarios of the model; in fact, the $s$~boson is much narrower than a corresponding heavy SM-like $h_{0}$~Higgs. The narrow-width approximation, therefore, remains valid and the aforementioned searches and projections apply throughout our heavy $s$ analyses.

At this stage, we have developed all the necessary tools in order to investigate the properties of the heavy $s$~boson by utilizing the LHC heavy Higgs searches \cite{CMSHeavyH,ATLASHeavyH} and projected 14~TeV exclusion limits \cite{ATLASHeavyS,Brownson:2013lka,CMSHeavyS} in the vector boson channels. In particular, employing the narrow-width approximation, these measurements and simulations constrain the $s$~production cross section times its branching ratio, as compared with that of a SM Higgs, $h_{0}$, with the same mass. As discussed in \cite{Chivukula:2013xka}, the amount of mixing between the SM $h_{0}$ Higgs and the gauge-singlet $s_{0}$ (c.f. \eqref{massbasis}) is constrained to small values, $\sin \chi \lesssim 0.3$, by the electroweak precision data and the properties of the discovered 125~GeV LHC $h$ signal. Furthermore, it was shown that, in this small mixing region, the $s$~production via gluon-fusion dominates over the other mechanisms. Hence, in order to adapt the LHC searches and exclusion projections in vector boson channels to study the $s$~boson, we may construct the following parametrization
\begin{align}
\mu(gg \to s\to VV)\equiv&\,\frac{\sigma(gg \to s) \times \textrm{BR}(s \to VV)}{\sigma(gg \to h_{0}) \times \textrm{BR}(h_{0} \to VV)} =\left|\frac{c_g^s}{\hat{c}_g^{s,\textrm{SM}}}\right|^2\frac{\textrm{BR}(s\to VV)}{\textrm{BR}(h_{0}\to VV)}\notag \\
=&\,\left|\frac{c_g^s}{\hat{c}_g^{s,\textrm{SM}}}\right|^2 \sin^2\chi  \left[ \frac{\Gamma_s^{\text{TOT}}(m_s)}{\Gamma_{h_{0}}(m_s)}\right]^{-1} \label{ggFsVV} \ ,
\end{align}
where we have inserted the appropriate coefficients from \eqref{heavyLeff}, and the ratio of the widths is given by \eqref{Ctots}.

In the previous treatment \cite{Chivukula:2013xka}, we have briefly discussed the constraints on the $s$~boson phenomenology, based on the LHC heavy Higgs searches in the vector boson channels for $\sqrt s = 7,8$~TeV \cite{CMSHeavyH,ATLASHeavyH}. In this section, we extend that discussion, by further considering the expected $s$~production cross section times branching ratio for $\sqrt s = 14$~TeV LHC, with an integrated luminosity of 300 and 3000 fb$^{-1}$. These simulations have been performed for the ATLAS detector assuming a SM-like heavy Higgs \cite{ATLASHeavyS}, and for the CMS detector assuming a heavy Higgs in the context of the 2HDM \cite{Brownson:2013lka,CMSHeavyS}. If the upcoming LHC run at $\sqrt s = 14$~TeV with the quoted luminosities does not detect a heavy Higgs-like scalar at or below the mass of 1~TeV, the projected exclusion limits will provide complementary constraints at 95\%~C.L. on the parameter space of the renormalizable coloron model.

To be specific, we utilize the heavy Higgs search and exclusion projections to understand the LHC sensitivity to the $\mu$ parameter (left-hand side of \eqref{ggFsVV}) in each case, and compare the latter with the theoretically derived expression (right-hand side of \eqref{ggFsVV}). If no heavy Higgs signal is detected by the LHC, this may be translated into 95\%~C.L. constraints on the model's variables. The $\sqrt s = 7,8$~TeV heavy Higgs searches \cite{CMSHeavyH,ATLASHeavyH} directly quote the $\mu$ parameter in various decay channels, and we use the strongest bounds determined by the $s$~decays into $WW$ and $ZZ$ final states. The $\sqrt s = 14$~TeV heavy Higgs exclusion projections \cite{ATLASHeavyS,Brownson:2013lka,CMSHeavyS}, however, quote the obtained production cross section times decay branching ratios, with the most stringent limits arising from the $ZZ$ decay channel. To construct the $\mu$ parameter for these simulations, we divide their obtained production cross section times $ZZ$ decay branching ratio by that of the SM $h_{0}$ Higgs, the latter being provided in \cite{ATLASHeavyS}.

\begin{figure}\begin{center}
\includegraphics[width=.5\textwidth]{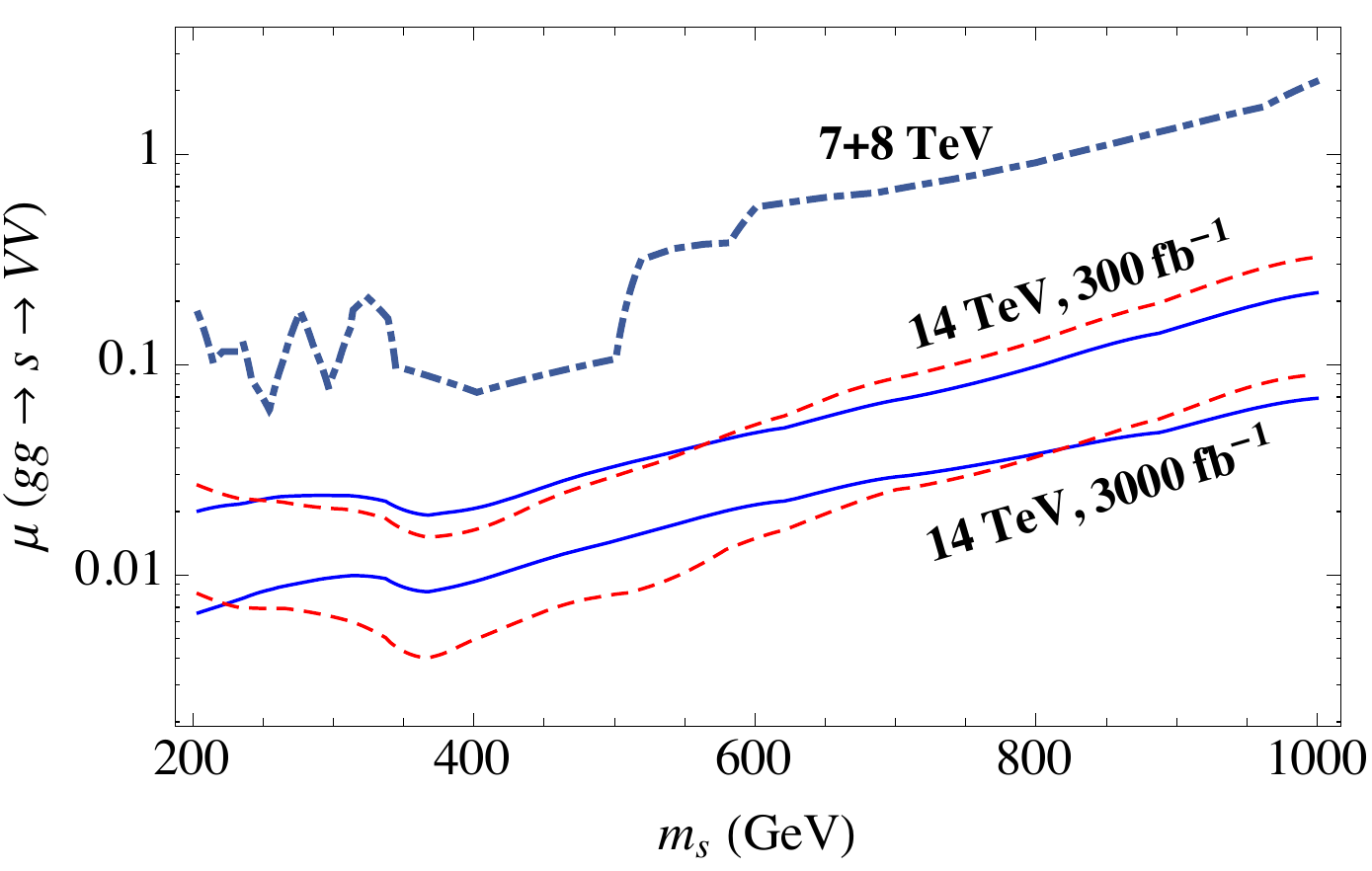}
\caption{The heavy $s$ Higgs production cross section times its vector boson decay branching ratio, divided by that of a corresponding SM $h_{0}$ Higgs with the same mass, as parameterized by $\mu$ in \eqref{ggFsVV}, for $200 \leq m_s \leq 1000$~GeV. The LHC $\sqrt s = 7,8$~TeV heavy Higgs exclusion limits at 95\%~C.L.  \cite{CMSHeavyH,ATLASHeavyH} are displayed by the dot-dashed line, with the most stringent bounds arising from the $WW$ and $ZZ$ final states. These data are complemented by the $\sqrt s = 14$~TeV projected exclusion limits at 95\%~C.L. with an integrated luminosity of 300 and 3000 fb$^{-1}$ for ATLAS (solid lines) assuming a SM-like heavy Higgs \cite{ATLASHeavyS}, and for CMS (dashed lines) assuming a heavy Higgs in the 2HDM \cite{Brownson:2013lka,CMSHeavyS}. In both 14~TeV projections, the strongest limits are determined by the $ZZ$ final states.}
\label{allmu}
\end{center}\end{figure}

Fig.~\ref{allmu} displays the most stringent values of the $\mu$ parameter as a function of the $s$ boson mass, incorporating the $\sqrt s = 7,8$~TeV LHC heavy Higgs search data ($WW$ and $ZZ$ final states) \cite{CMSHeavyH,ATLASHeavyH}, as well as the exclusion projections for $\sqrt s = 14$~TeV with an integrated luminosity of 300 and 3000~fb$^{-1}$ ($ZZ$ final states), based on search projections for a SM-like \cite{ATLASHeavyS} and a 2HDM-like \cite{Brownson:2013lka,CMSHeavyS} heavy Higgs. Clearly, future searches with higher center-of-mass energies and luminosities are anticipated to yield more stringent upper bounds on the $\mu$ parameter---and the model's parameter space---at 95\%~C.L., if no heavy Higgs-like scalar is detected at the LHC in the displayed mass range. Nonetheless, one notes the similarity in the obtained results for the SM-like and the 2HDM-like heavy Higgs projections. The differences between the two may be attributed to the differences in widths of the two particles. Given this similarity, in the following, we shall apply only the SM-like Higgs expected $gg\to s \to ZZ$ projections \cite{ATLASHeavyS} to our analyses of the $s$~boson phenomenology.

Separately, we consider the projected ability of the ATLAS experiment at $\sqrt s = 14$~TeV ATLAS with 300~fb$^{-1}$ of integrated luminosity to more precisely measure the properties of the existing 125~GeV $h$~Higgs \cite{ATLASHproj}. Similar to the case of the heavy $s$~scalar, these $h$~boson measurements present additional constraints on the parameter space of the model, assuming no large deviations from the current measurements of the $h$ properties are discovered at the $\sqrt s = 14$~TeV LHC run. The strongest bounds in this study arise from the gluon-fusion production mechanism and the $ZZ$~decay mode; hence, we take only these particular channels into account. The analogous expression for the $\mu$~parameter \eqref{ggFsVV} of the $h$~Higgs has been provided in \cite{Chivukula:2013xka}; it formally depends on all the input parameters \eqref{freepar}, except for the mass of the $s$~boson, $m_{s}$. According to \cite{ATLASHproj}, at 300~fb$^{-1}$, the experimental relative uncertainty of the signal strength is $\Delta\mu/\mu=0.06$.\footnote{If the current theoretical uncertainty is included, the total uncertainty becomes $\Delta \mu/\mu = 0.13$. The theoretical uncertainties, however, are expected to decrease---and therefore, in this paper, we illustrate the sensitivity of the 14~TeV LHC to this model using the experimental uncertainties alone.} Assuming $\mu = 1$, and requiring no exclusion by the previous $\sqrt s =7,8$~TeV best fit at 95\%~C.L. \cite{Chivukula:2013xka}, we impose the 95\%~C.L. exclusion constraint from the $\sqrt s =14$~TeV $h$~projections at 300~fb$^{-1}$ to find the sensitivity of high-energy LHC data to this model.

\begin{figure}
\includegraphics[width=.49\textwidth]{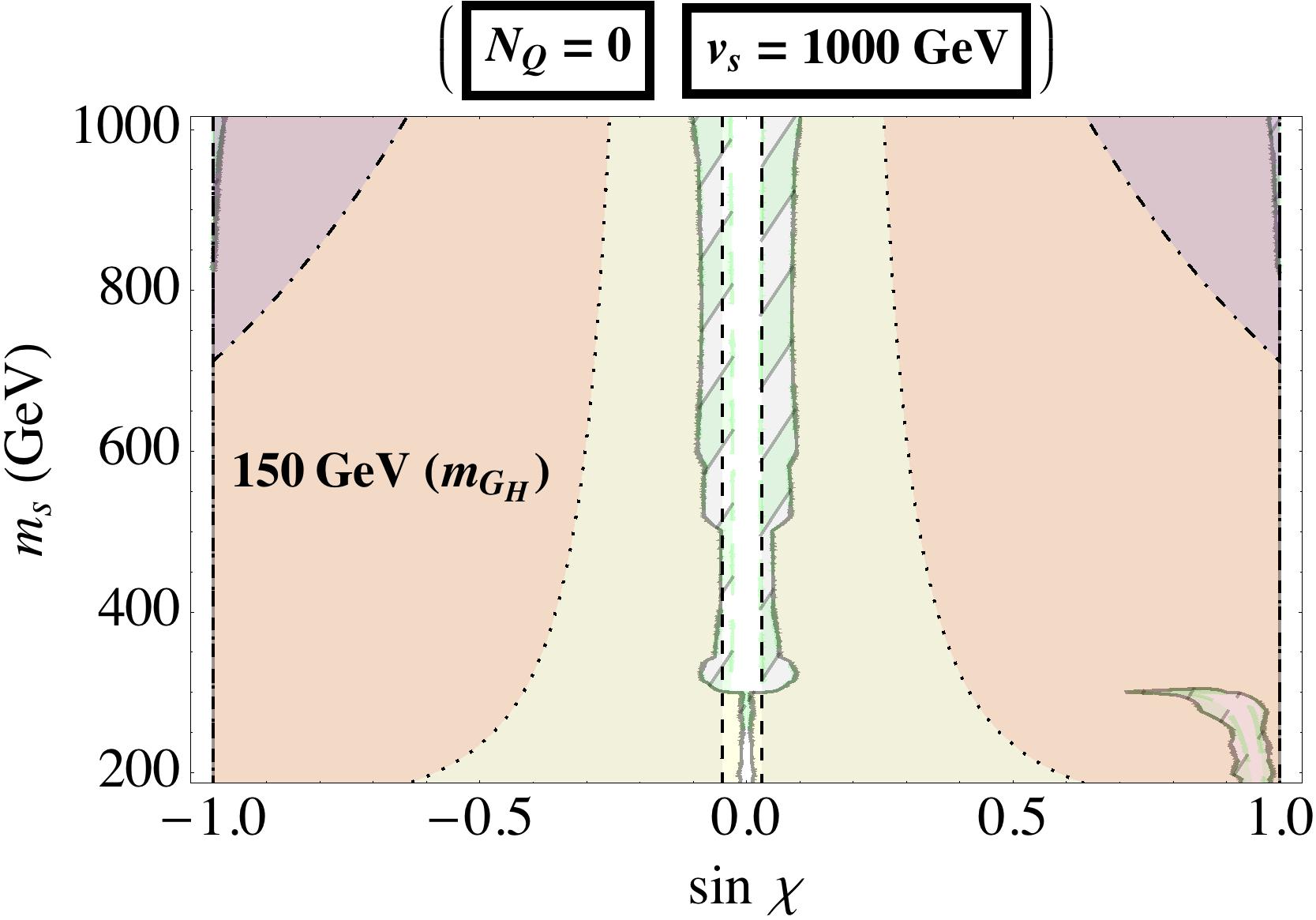}
\includegraphics[width=.49\textwidth]{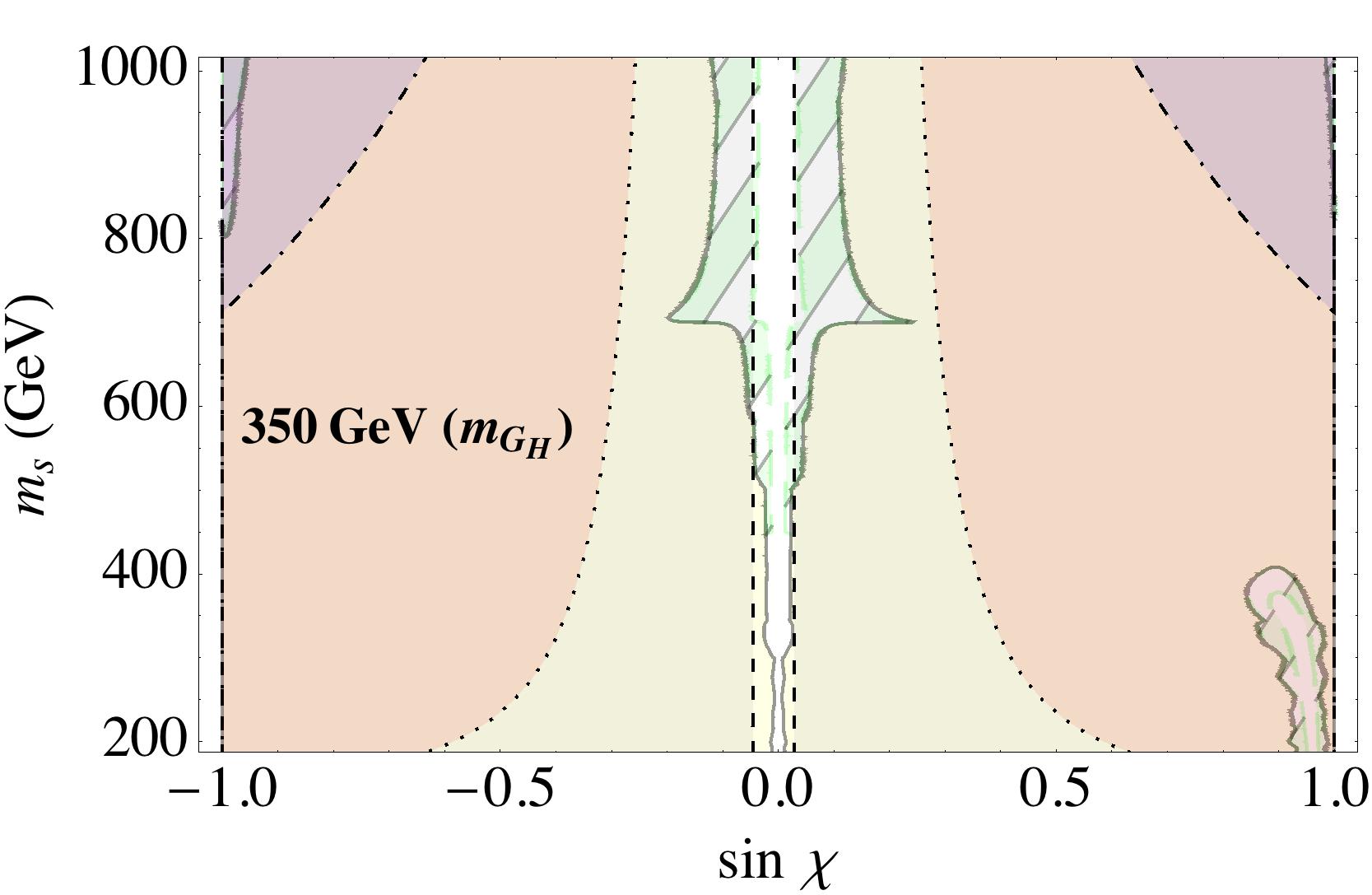}
\includegraphics[width=.49\textwidth]{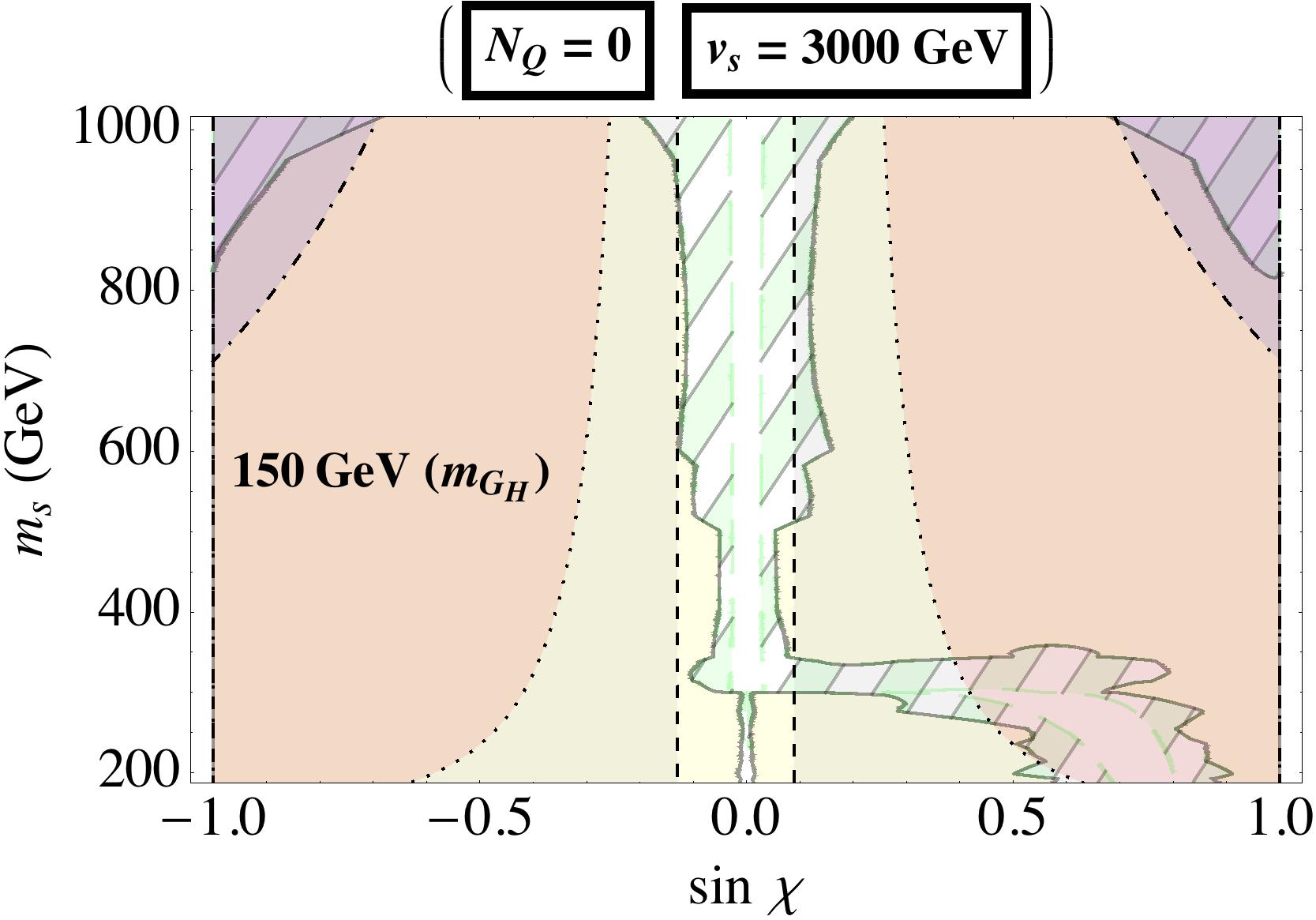}
\includegraphics[width=.49\textwidth]{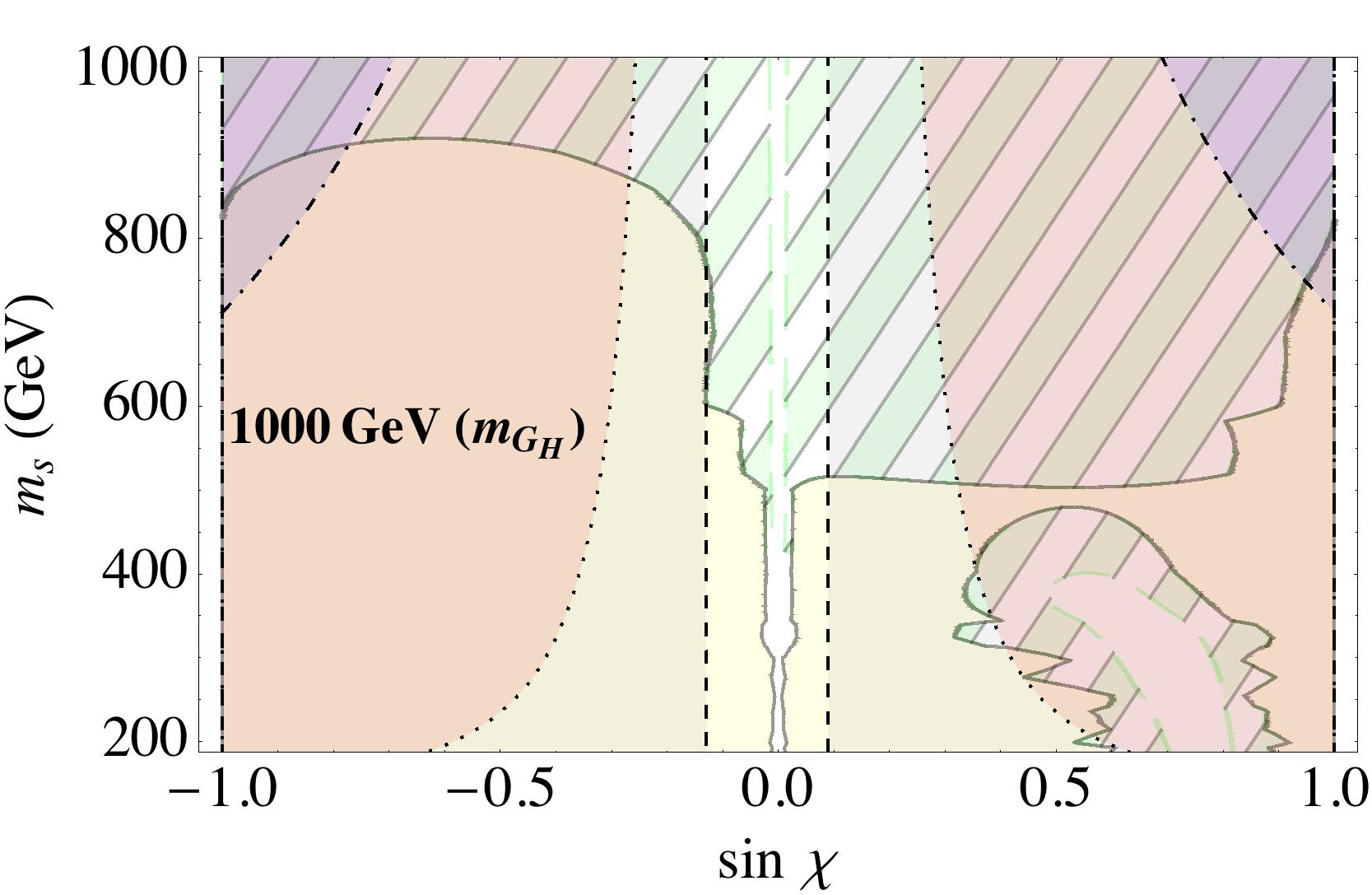}
\includegraphics[width=.49\textwidth]{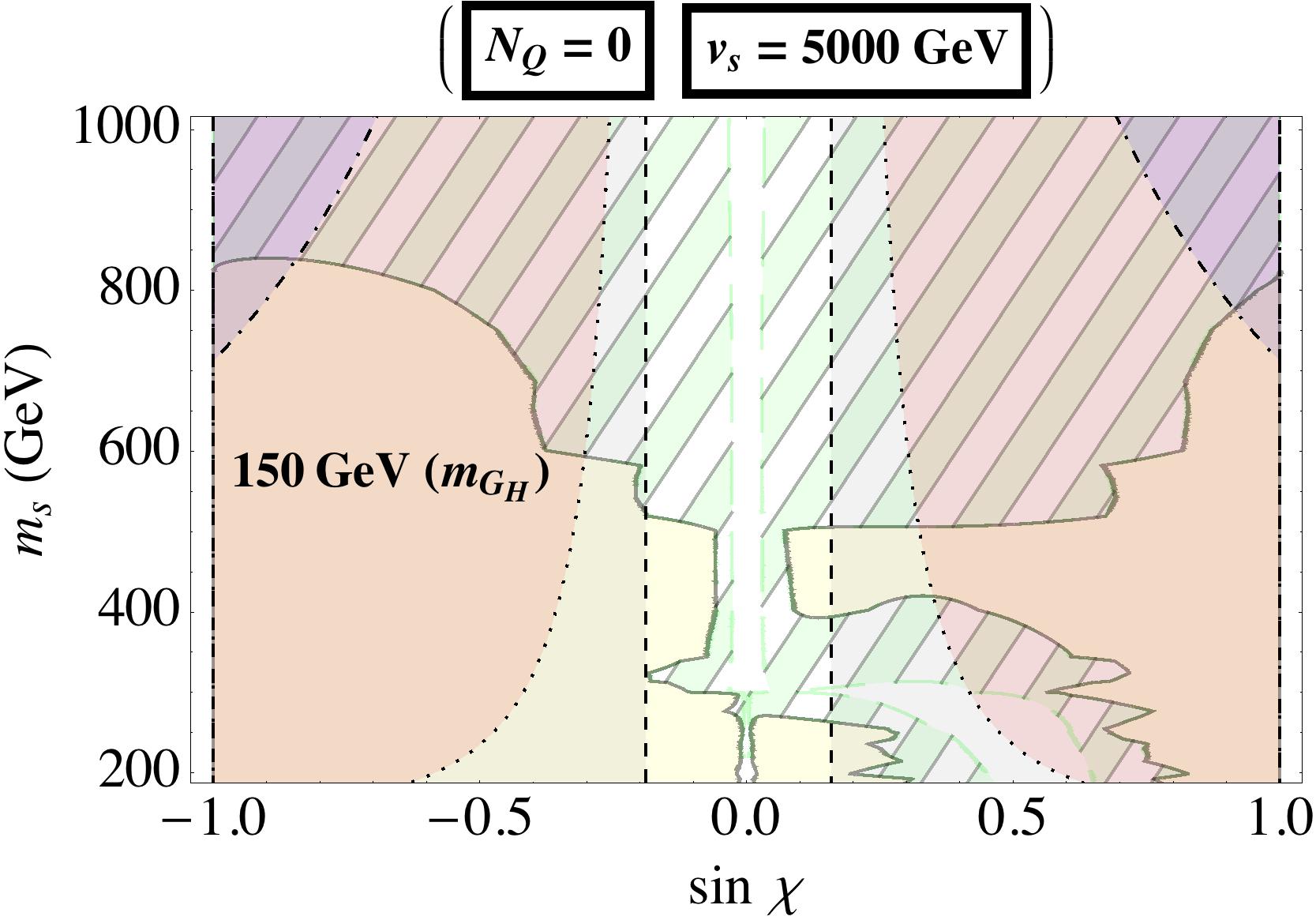}
\includegraphics[width=.49\textwidth]{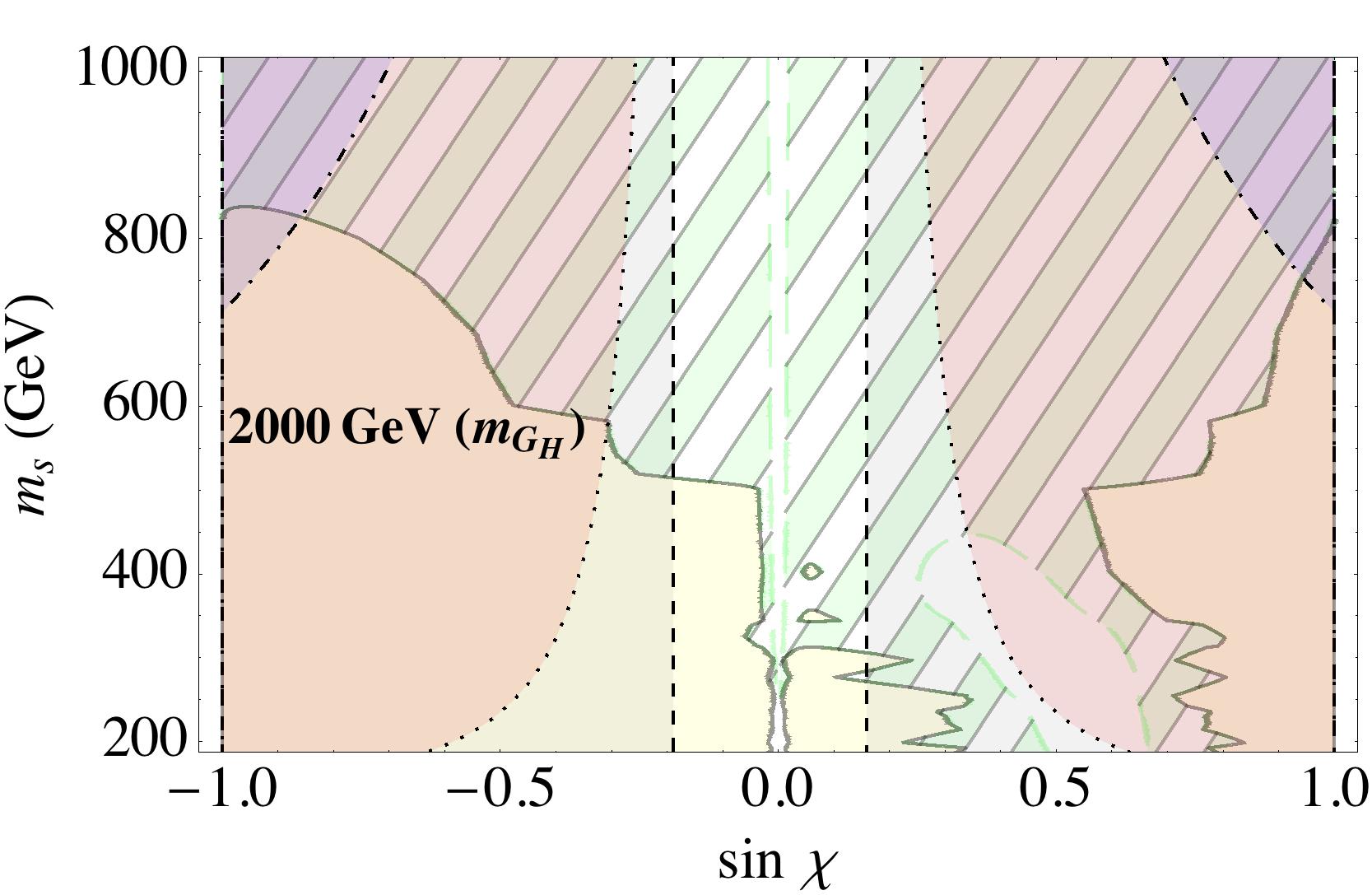}
\caption{95\%~C.L. exclusion contours for the scenario with no spectator fermion generation, $N_{Q}=0$, represented in the $m_s - \sin\chi$ plane, covering the heavy $s$~scalar mass range $200 \leq m_s \leq 1000$~GeV and the full range of mixing angle values $-1\leq \sin \chi \leq 1$. A universal pseudoscalar mass, $m_\mathcal{A} = 150$~GeV, has been used for the purpose of illustration. Three selected values of the singlet VEV, $v_s$, are displayed in the three rows, within which the scalar color-octet mass, $m_{G_H}$, is varied from light (left plot) to heavy (right plot). The exclusion limits arise from imposing unitarity (dot-dashed purple), electroweak precision tests (dotted orange), LHC direct measurements of the 125~GeV $h$~Higgs (vertical dashed gray), and the $\sqrt s = 7,8$~TeV LHC heavy $s$~boson searches (solid yellow). The inclined-shaded green region corresponds to the heavy $s$~boson exclusion projections for $\sqrt s = 14$~TeV ATLAS with an integrated luminosity of 300 fb$^{-1}$. Given their heavy nature, the dependence on the coloron and spectator masses, $M_C$ and $M_Q$, is negligible throughout.}
\label{resultNQ0}
\end{figure}

\begin{figure}
\includegraphics[width=.49\textwidth]{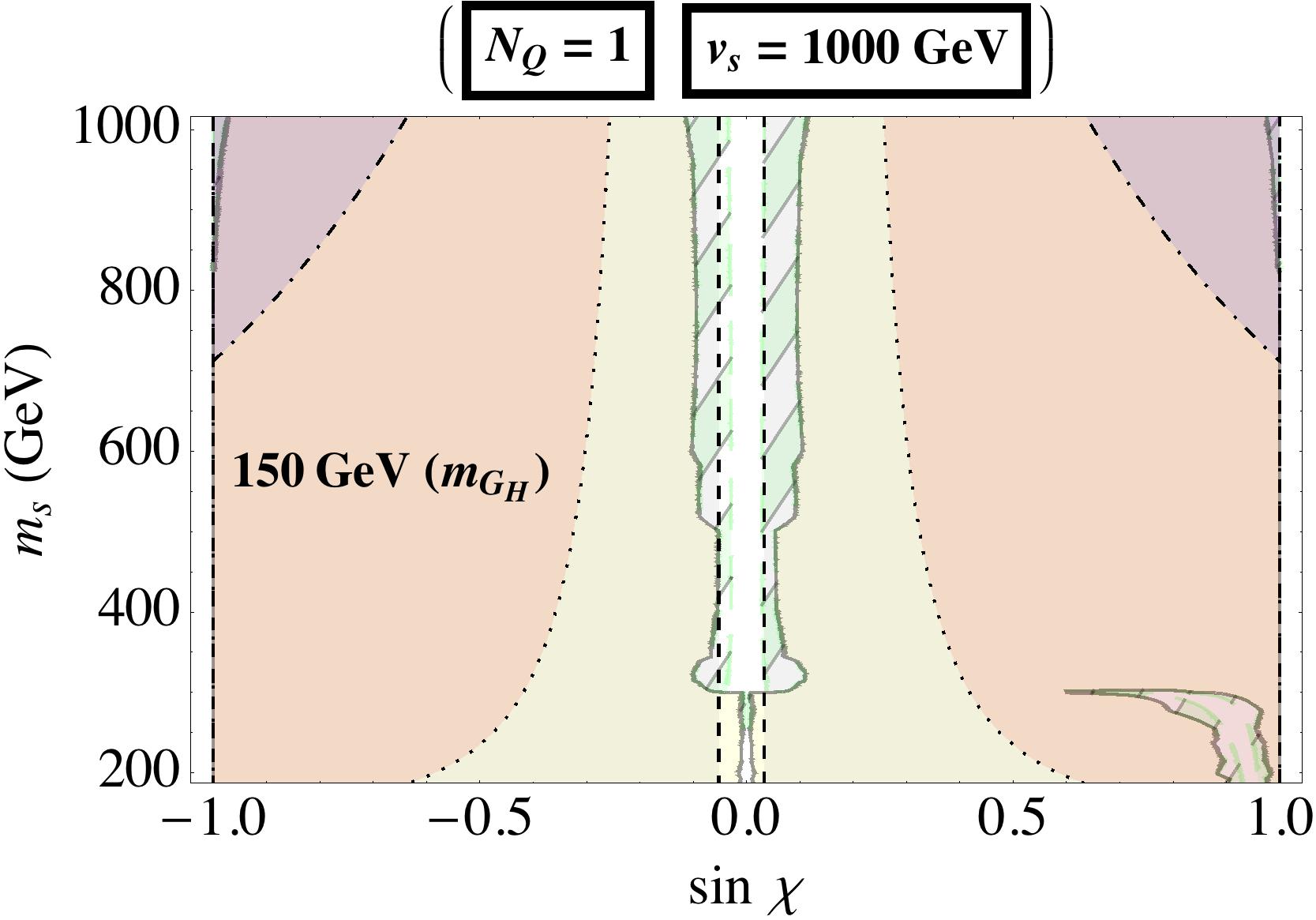}
\includegraphics[width=.49\textwidth]{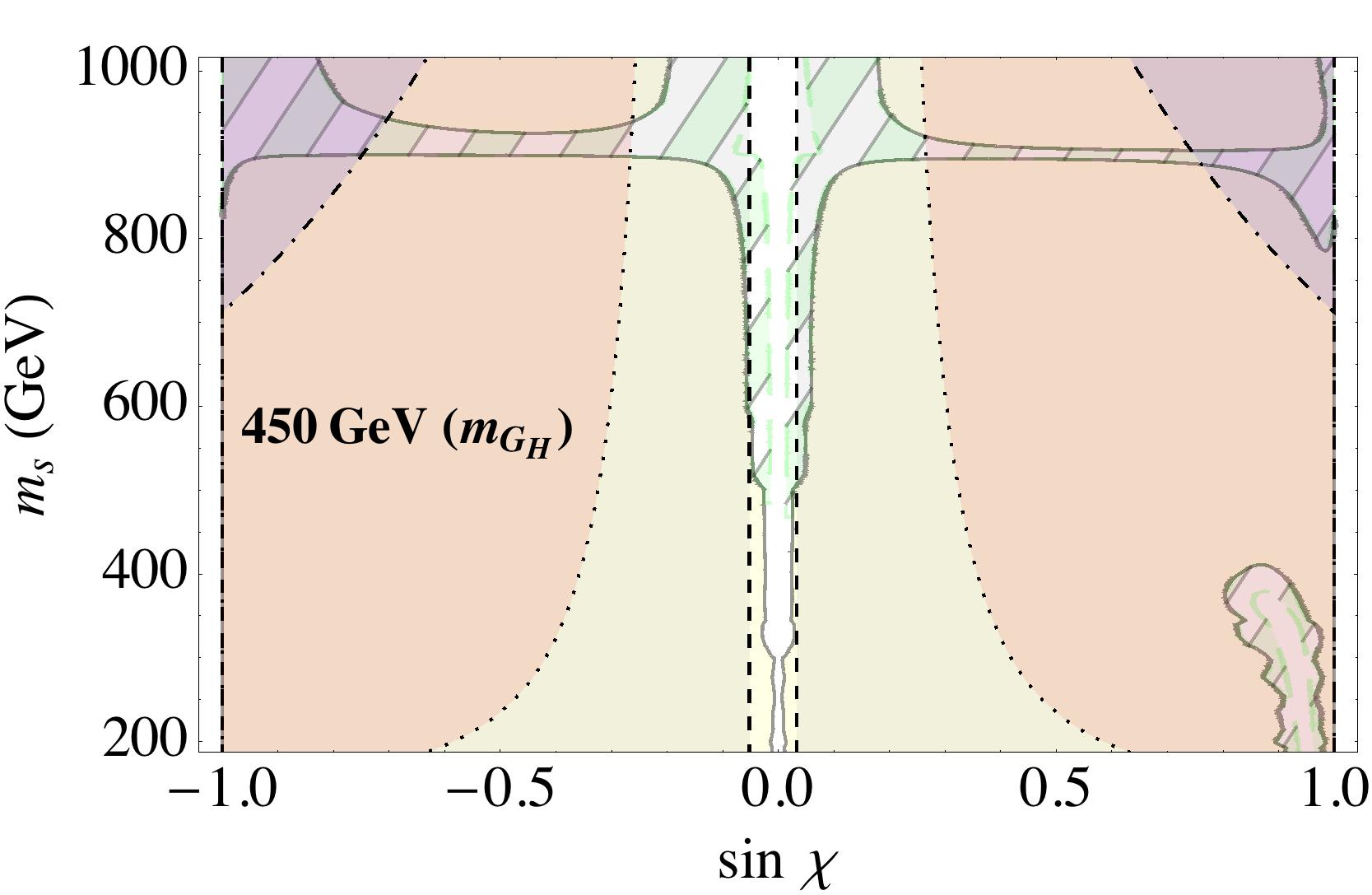}
\includegraphics[width=.49\textwidth]{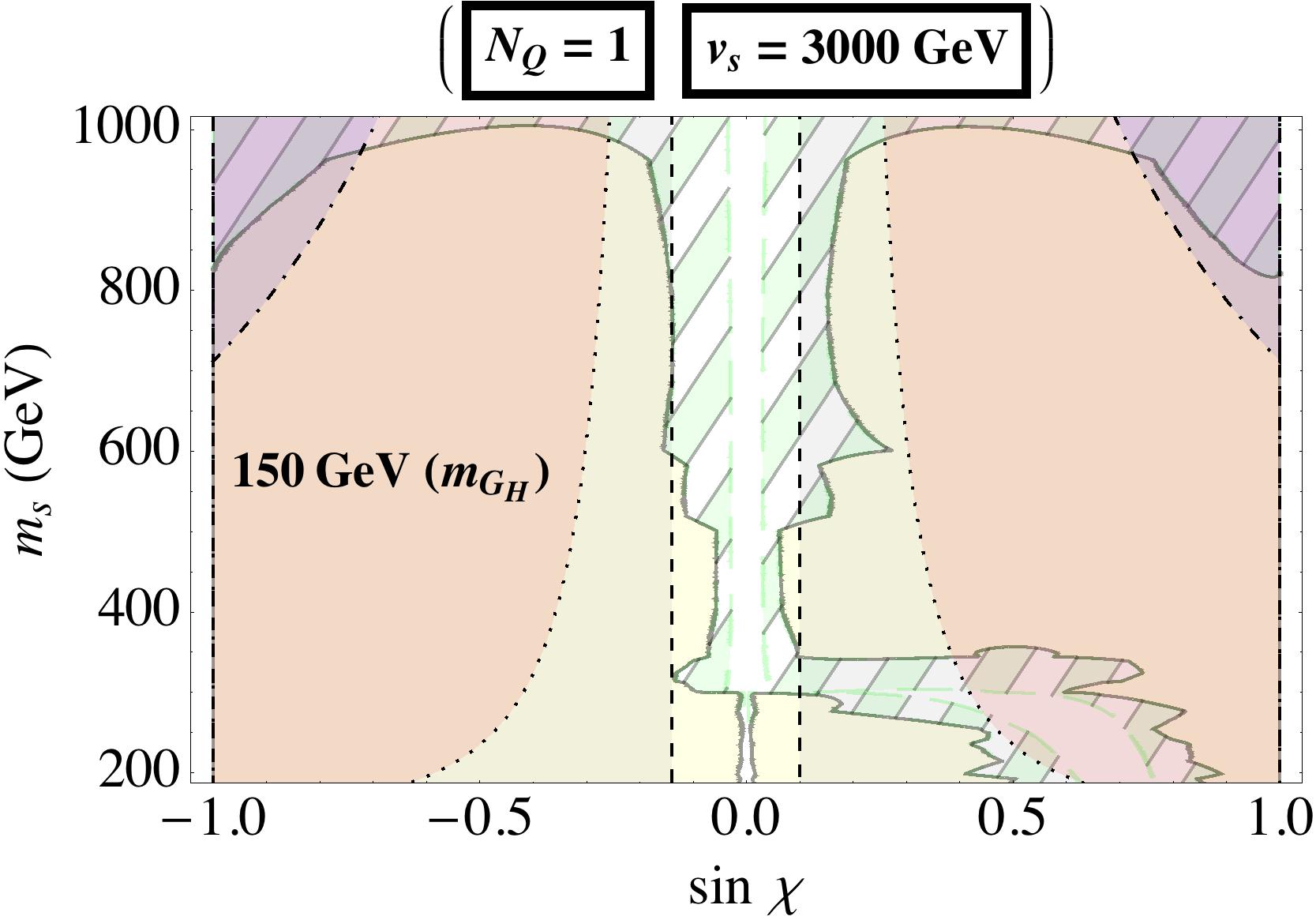}
\includegraphics[width=.49\textwidth]{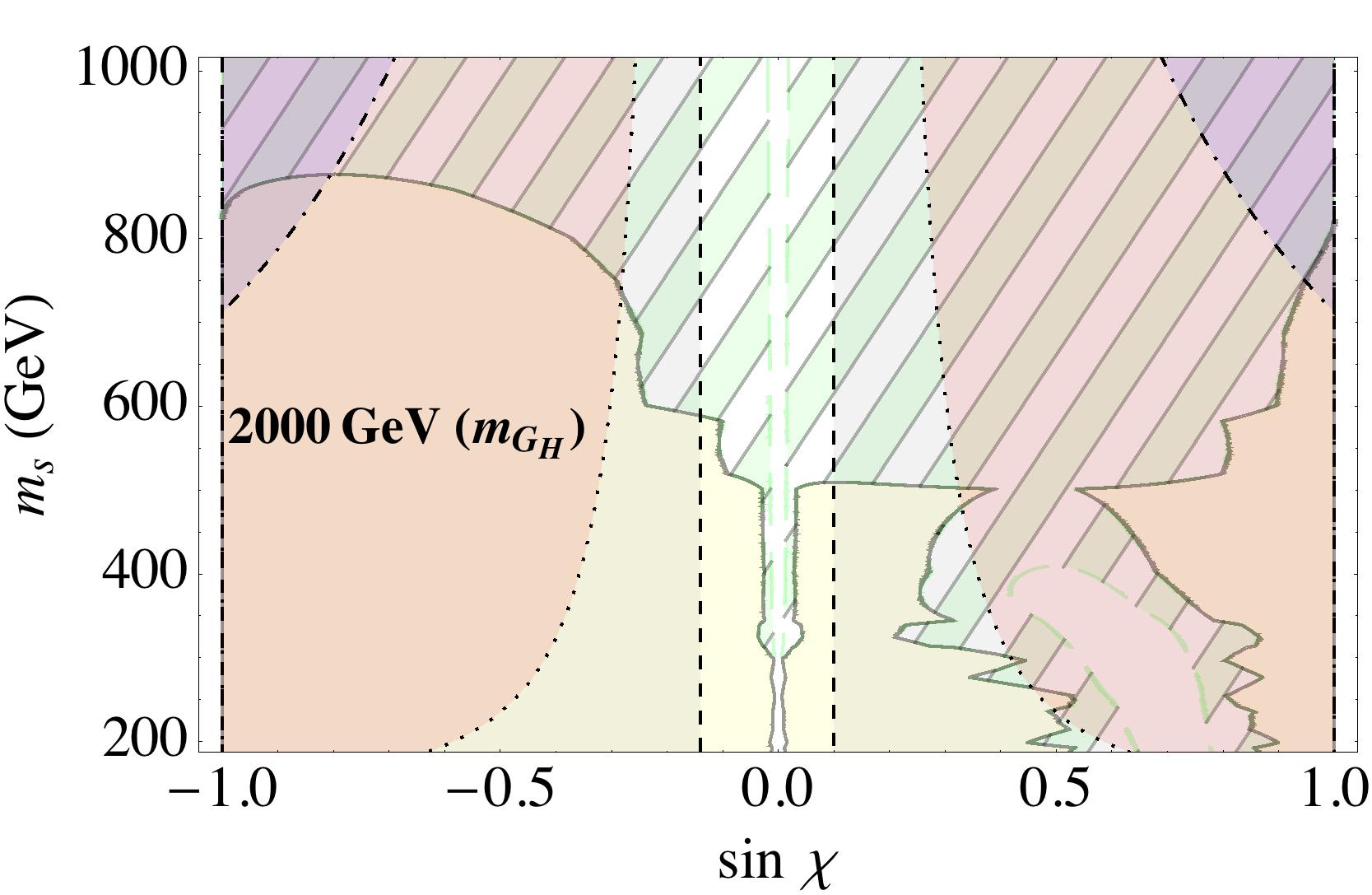}
\includegraphics[width=.49\textwidth]{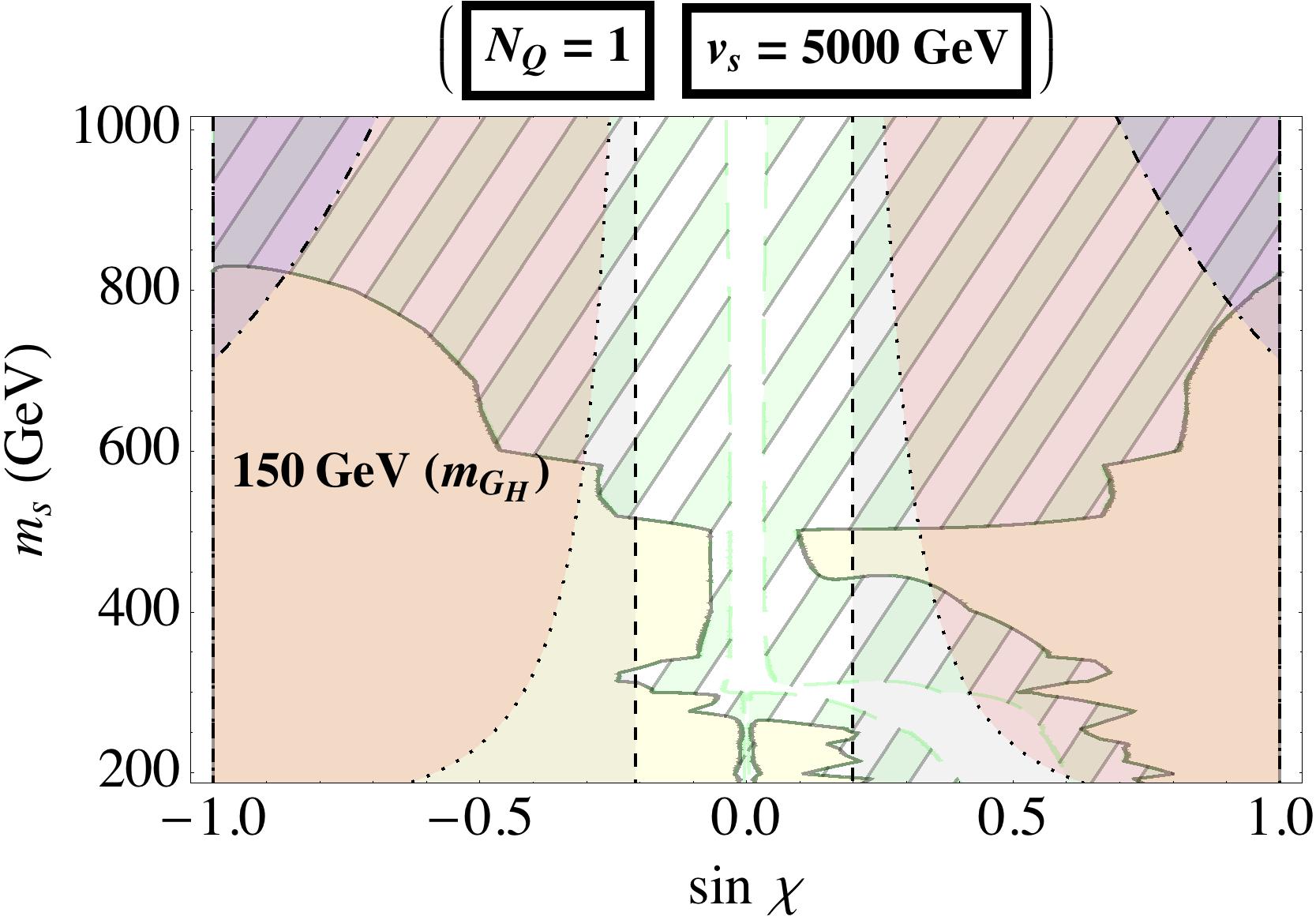}
\includegraphics[width=.49\textwidth]{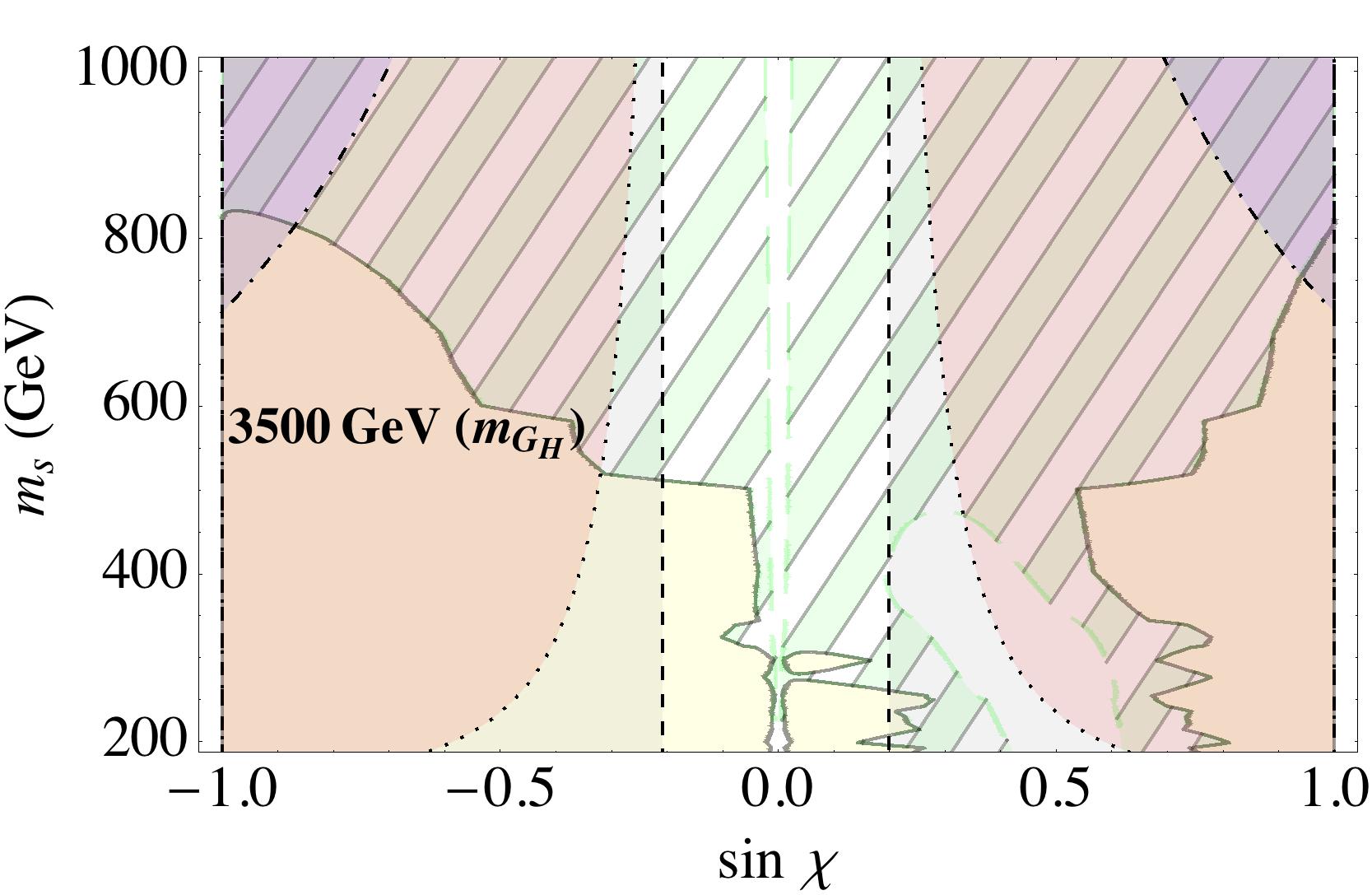}
\caption{95\%~C.L. exclusion contours for the scenario with one spectator fermion generation, $N_{Q}=1$, represented in the $m_s - \sin\chi$ plane, covering the heavy $s$~scalar mass range $200 \leq m_s \leq 1000$~GeV and the full range of mixing angle values $-1\leq \sin \chi \leq 1$. (For details, see the caption of Fig.~\ref{resultNQ0})}
\label{resultNQ1}
\end{figure}

\begin{figure}
\includegraphics[width=.49\textwidth]{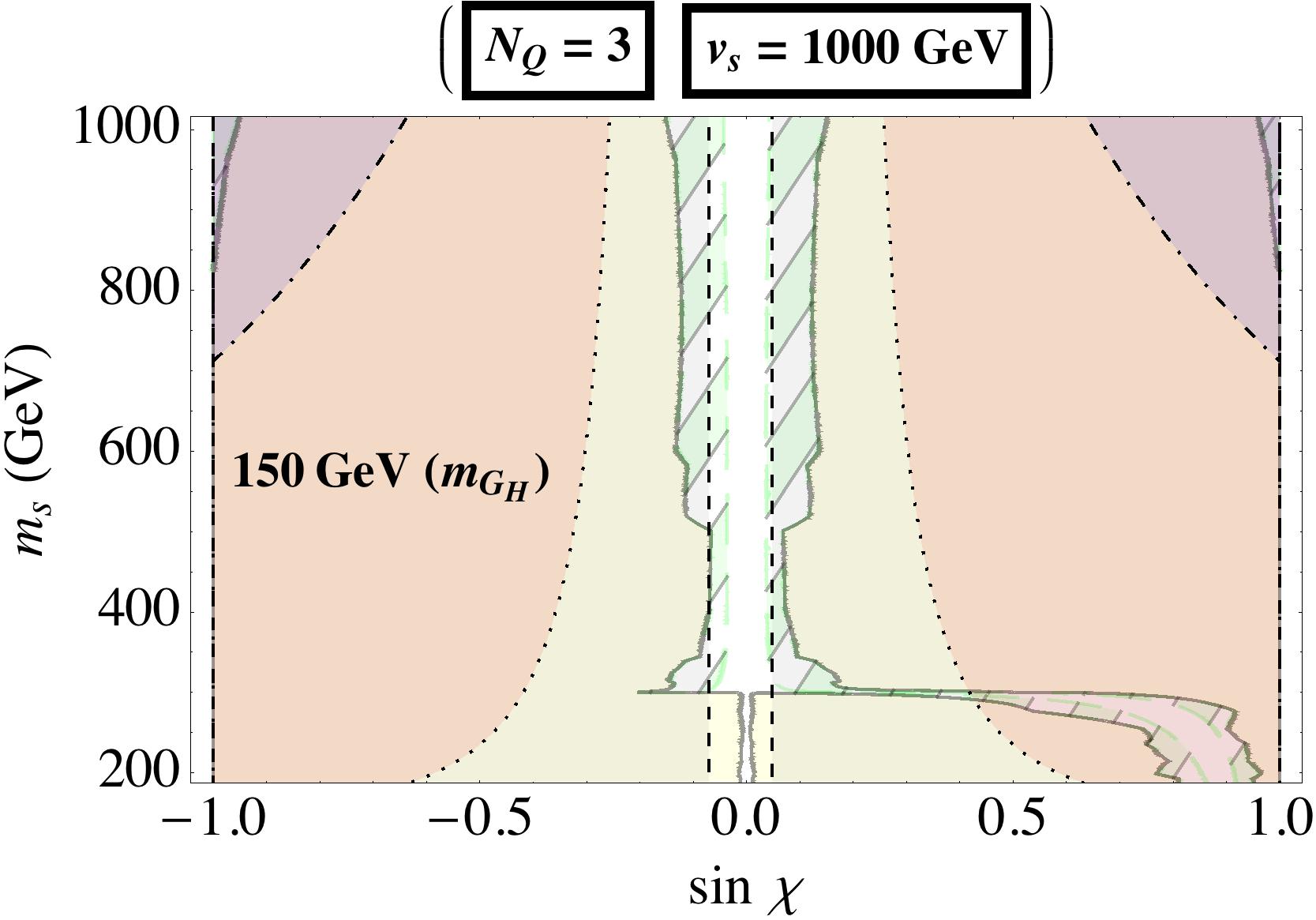}
\includegraphics[width=.49\textwidth]{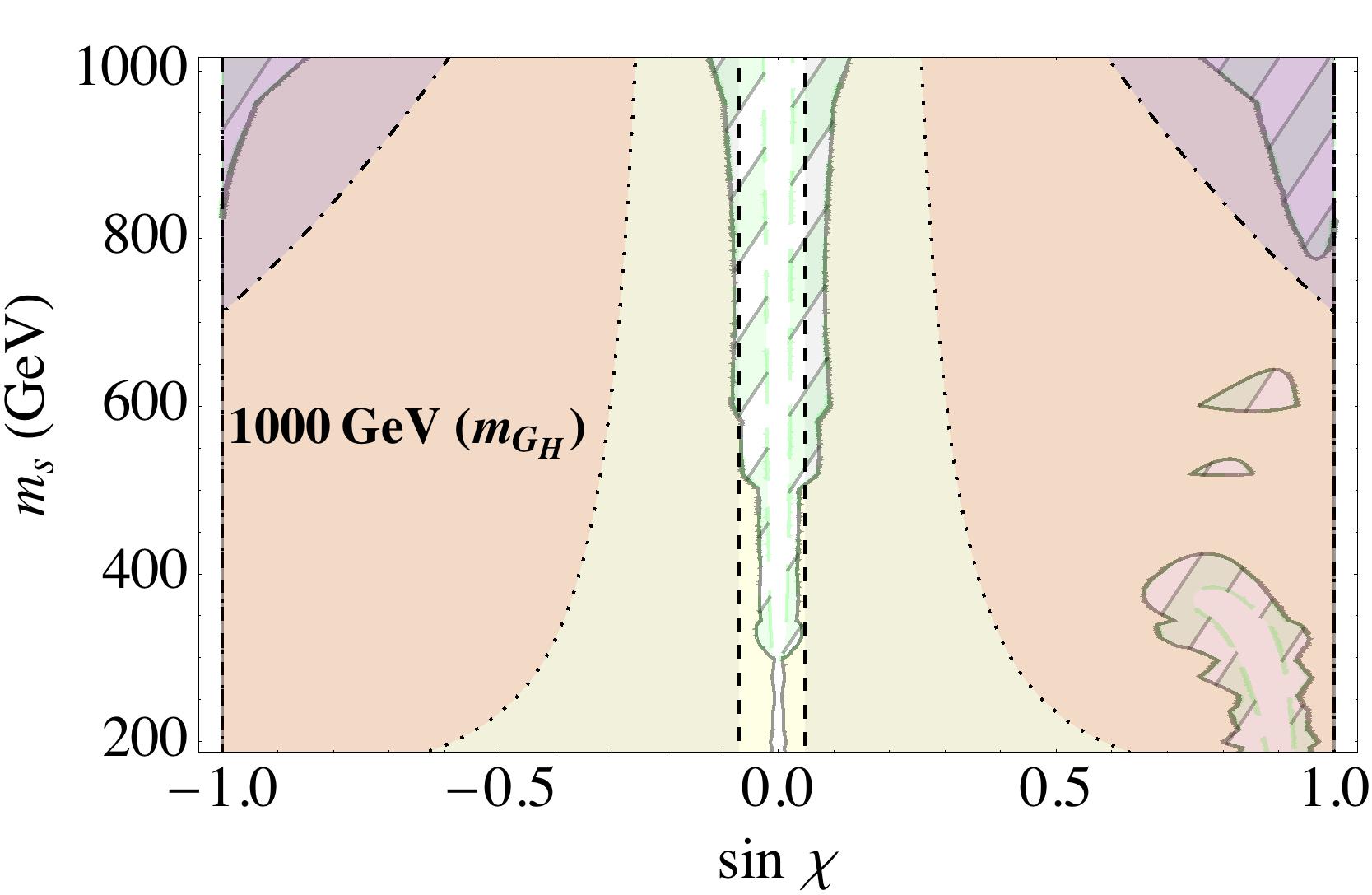}
\includegraphics[width=.49\textwidth]{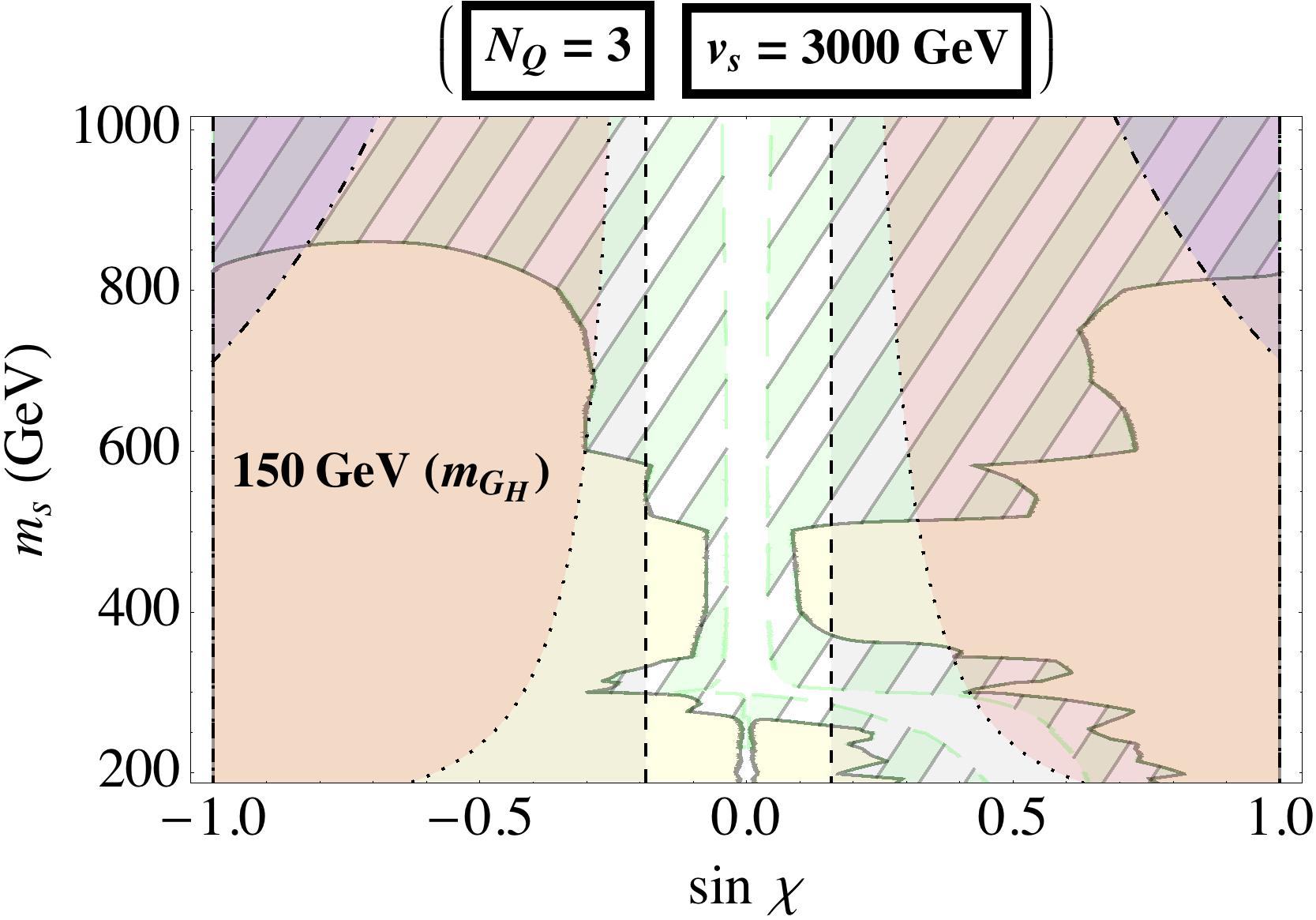}
\includegraphics[width=.49\textwidth]{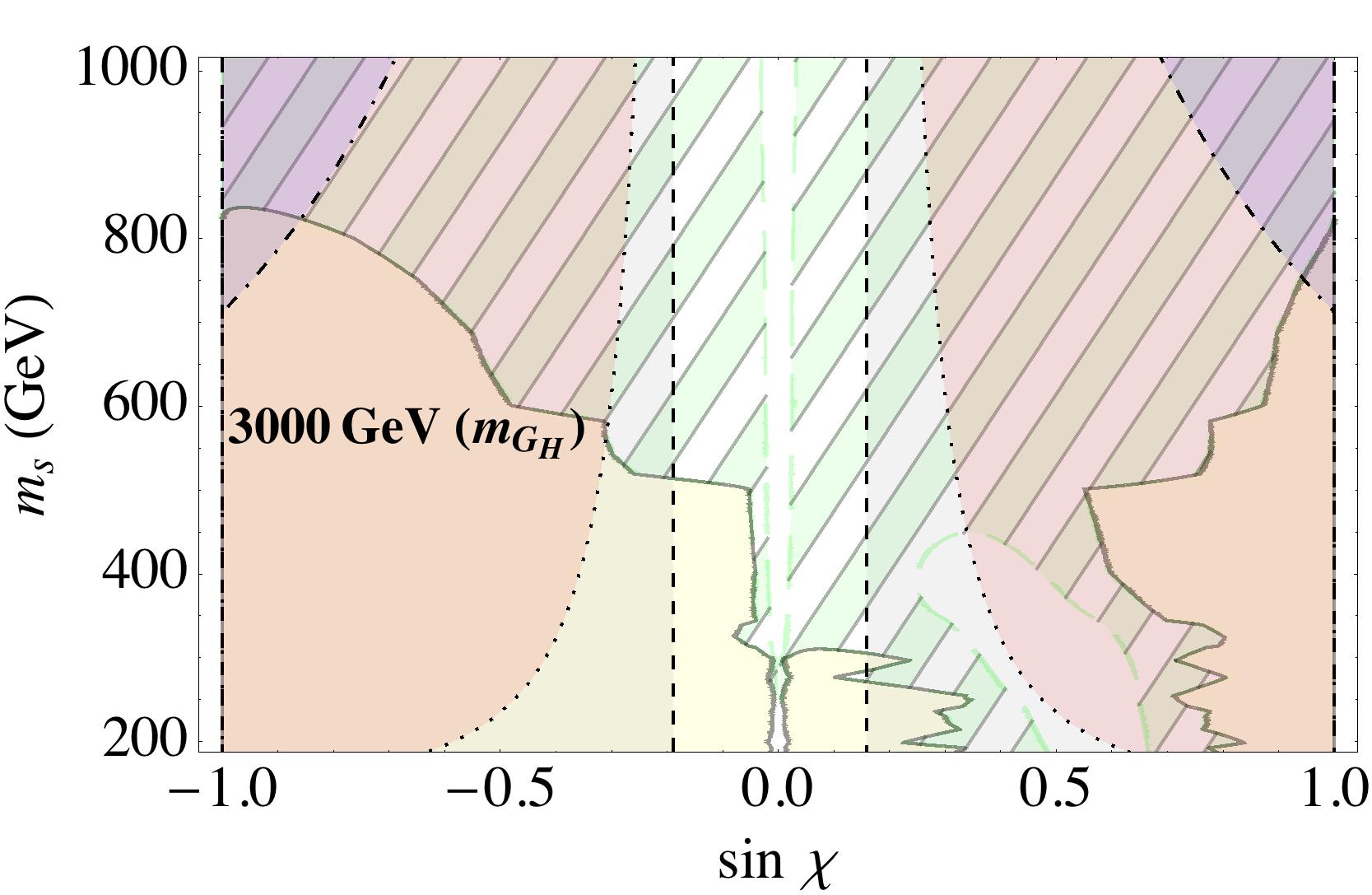}
\includegraphics[width=.49\textwidth]{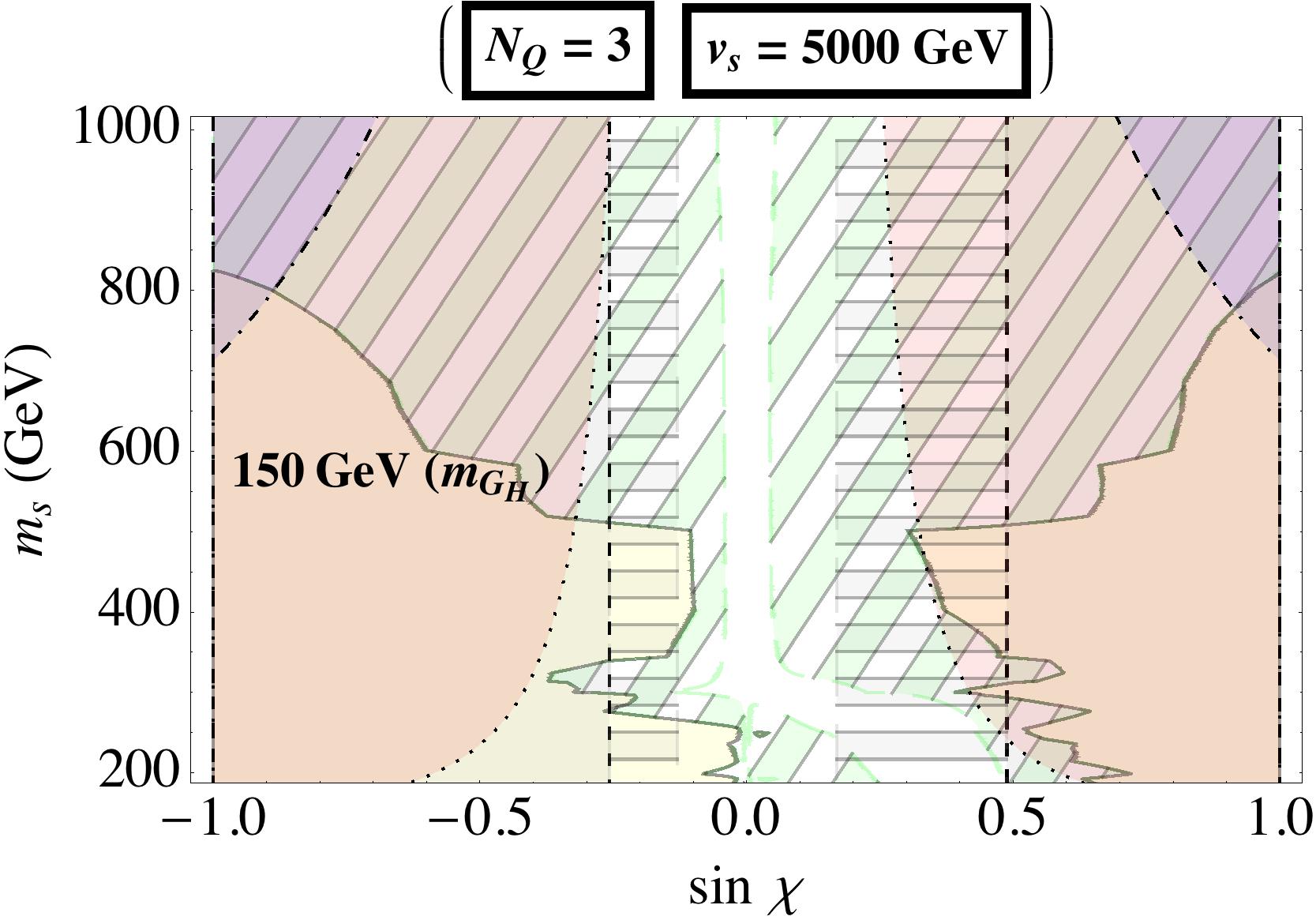}
\includegraphics[width=.49\textwidth]{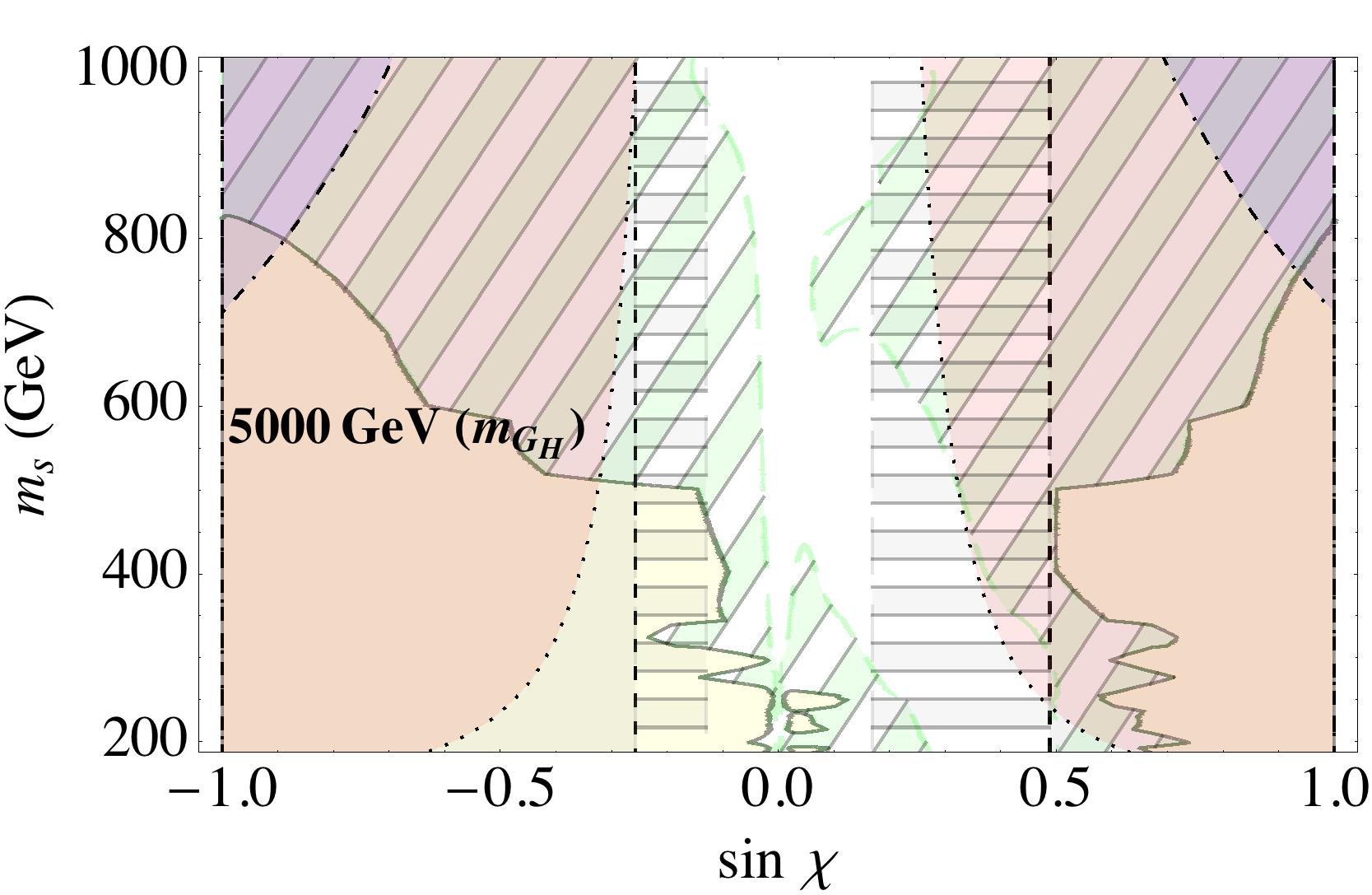}
\caption{95\%~C.L. exclusion contours for the scenario with three spectator fermion generations, $N_{Q}=3$, represented in the $m_s - \sin\chi$ plane, covering the heavy $s$~scalar mass range $200 \leq m_s \leq 1000$~GeV and the full range of mixing angle values $-1\leq \sin \chi \leq 1$. The additional horizontally-shaded gray region in the $v_{s}=5$~TeV case accounts for the exclusion by more precise measurements of the 125~GeV $h$~Higgs. (For details, see the caption of Fig.~\ref{resultNQ0})}
\label{resultNQ3}
\end{figure}

As discussed in section~\ref{review}, we present our plots for three scenarios containing various spectator fermion generations, $N_Q=0$, 1, and 3. The results are exhibited, respectively, in Figs.~\ref{resultNQ0}-\ref{resultNQ3}. The panels incorporate the constraints explored in \cite{Chivukula:2013xka} arising from unitarity, electroweak precision tests, and LHC 125~GeV $h$~Higgs data, as well as the LHC heavy Higgs searches with $\sqrt s = 7,8$~TeV \cite{CMSHeavyH,ATLASHeavyH}, and the $s$~and~$h$~Higgs ATLAS projections for $\sqrt s = 14$~TeV with an integrated luminosity of 300~fb$^{-1}$ at 95\%~C.L. \cite{ATLASHeavyS,ATLASHproj}.  Highlighting the importance of the sign associated with $\sin \chi$ for the LHC analyses (see the discussion below \eqref{formfact}), the current plots are extended to cover the entire range $-1\leq \sin \chi \leq 1$.\footnote{Note that the tree-level studies of the stability, unitarity, and the electroweak precision tests are insensitive to this sign \cite{Chivukula:2013xka}; only the LHC analyses of the 125~GeV $h$ Higgs and the heavy $s$ Higgs depend on it.} For each scenario, three benchmark values of the singlet VEV, $v_s$, have been selected, within which the mass of the color-octet scalar boson, $m_{G_H}$, is varied from light to heavy. A universal pseudoscalar mass, $m_{\mathcal A} = 150$~GeV, is also assumed.\footnote{As explained in \cite{Chivukula:2013xka}, the stability constraint becomes relevant in the displayed $m_{s} - \sin \chi$ region only for $m_{\mathcal A} \gtrsim 400$~GeV, imposing a lower bound on $m_{s}$ (c.f. \eqref{masscond}). Hence, for the quoted pseudoscalar masses, this theoretical constraint is absent in our displayed figures.} We have, furthermore, set the spectator and coloron TeV-range masses to $M_Q=1$~TeV and $M_C=3$~TeV throughout, as reference values. For colorons and spectators this heavy, the explicit dependence of the analysis on the precise values of $M_C$ and $M_Q$ is negligible. The limits arising from more precise measurements of the 125~GeV $h$~boson properties are weaker or at most comparable to those from the $s$~scalar projections in all depicted scenarios, except for $N_Q=3$ with very high singlet VEV and color-octet masses. The $h$~Higgs projection constraints are, therefore, omitted in the remaining scenarios.
 
\section{Discussion of the Results}\label{disc}

Within each $N_{Q}$ scenario, it is evident that a larger singlet VEV, $v_s$, enlarges the viable parameter space by alleviating the exclusion constraints, and, at the same time, accommodates heavier scalar color-octets. Moreover, for the different signs of the mixing angle, $\sin \chi$, one observes an asymmetry in all LHC $h$ and $s$  searches and projections, indicating the sensitivity of these analyses to this sign. The origin of this sensitivity may be traced back to the constructive or destructive role played by the sign in the heavy $s$ (and $h$) production via gluon fusion (see the discussion below \eqref{formfact}). It is, furthermore, interesting to note that the pseudoscalar mass, $m_{\mathcal A}$, has the effect of determining the lower bound on the mass of the heavy $s$~scalar near the $\sin \chi \simeq 0$ region.\footnote{We note that this observation is in accordance with a similar conclusion deduced from the stability condition, as discussed in \cite{Chivukula:2013xka} (see also \eqref{masscond}), but we find here a much stronger constraint.} This is due to the fact that the $s$~decay channel into a pair of pseudoscalars is proportional to $\cos\chi$ and becomes accessible at a lower $m_{s}$ for a lighter $\mathcal A$, further suppressing its decay branching ratios into pairs of vector bosons. As anticipated, the projections demonstrate that the $\sqrt s = 14$~TeV LHC is far more sensitive to a potential heavy scalar signal than the $\sqrt s = 7,8$~TeV searches already carried out. In fact, the heavy $s$~boson projections with 300~ft$^{-1}$ luminosity already cover practically the entire remaining region of the parameter space in all possible scenarios, except for the very high singlet VEV and color-octet masses (deep in the TeV region) that are only available within the $N_Q=3$ scenario. This signals the interesting predictive power of the model, which can be thoroughly probed by the imminent LHC searches. Given this increased sensitivity, in case no heavy Higgs signal is detected by the LHC at $\sqrt s = 14$~TeV, the displayed projections at 300~ft$^{-1}$ tightly constrain at 95\%~C.L. the model's parameter space, while the 3000~ft$^{-1}$ luminosity data (not shown) completely rule out the model in the depicted region of interest (c.f. Fig.~\ref{allmu}).

A comparison of the three models with different spectator generation contents, as presented in Figs.~\ref{resultNQ0}-\ref{resultNQ3}, reveals that the model with three spectator fermion generation ($N_Q=3$) is the scenario that is least constrained by the current theoretical and experimental exclusions for a wide range of the free parameters, such as the scalar color-octet mass and the singlet VEV. This is attributable to the large cancellation occurring between the coloron and spectator contributions in the effective $s$~coupling to the gluon pair \eqref{delcgs}. The projected bounds on the properties of the $h$~Higgs, on the other hand, do not suffer from such cancellations in this region of parameter space and become relevant, providing additional potential constraints at 300~ft$^{-1}$. Consequently, in all three cases, it is generally true that nearly the entire parameter space of the models lies within reach of the $\sqrt{s} = 14$~TeV LHC.

Finally, let us elaborate on how a potential discovery of a heavy scalar boson by the LHC may be identified with the $s$~boson of the minimally extended color sector. As discussed in Section~\ref{review}, within the context of the renormalizable coloron model, the heavy $s$~boson is necessarily accompanied by a multitude of additional scalar, vector, and fermionic degrees of freedom.  Being color-charged many of these additional states may reveal their existence in a high center-of-mass energy hadron collider. A heavy $s$~scalar---in contrast with the scalar in many other proposed ``Higgs portal'' models \cite{Schabinger:2005ei,Barbieri:2005ri,Patt:2006fw}---is likely to be discovered in association with other (colored) resonances of comparable mass. Moreover, the existence of these additional colored states may enhance the effective gluon coupling to the heavy $s$~boson \eqref{eq:cgs}, potentially compensating for the suppressed coefficient proportional to the small mixing angle ($\sin \chi$), which is favored by the current experimental constraints.\footnote{The enhanced gluon coupling makes vector boson fusion production irrelevant for the $s$-boson, another potential distinguishing feature of this state.} This may, in turn, enhance the heavy Higgs production cross section---an effect absent in the color-singlet Higgs portal extensions---and potentially visible in the global fitting analyses of the data.

\section{Conclusion}\label{concl}

We have studied the collider phenomenology of a heavy color-singlet scalar boson originating in models with a minimally extended color sector. The properties of the heavy scalar---a color-singlet with some degree of weak interactions---turn out to be closely related to the properties of additional (colored) states in the theory, which may include scalars, vectors, or fermionics.

Incorporating the projected $\sqrt{s}=14$~TeV 95\%~C.L. exclusion limits, along with the previously obtained exclusion bounds \cite{Chivukula:2013xka} imposed by unitarity, electroweak precision tests, LHC 125~GeV $h$~Higgs searches, and $\sqrt s = 7,8$~TeV LHC heavy Higgs searches, we have investigated various scenarios within the renormalizable coloron model, with zero, one, or three generations of the spectator fermions. We have demonstrated the sensitivity of the future high center-of-mass energy and luminosity LHC searches to this model. Our results are summarized in Figs.~\ref{resultNQ0}-\ref{resultNQ3}, covering the free parameter space of the theory for the scenarios with different numbers of spectator fermion generations.

We find that the upcoming LHC searches should be sensitive to an $s$~scalar of mass less than 1~TeV for essentially all of the model parameter space in which the $h$~state differs from the Higgs boson of the SM.  More precisely, unless the mixing angle, $\sin\chi$, is zero, the 14~TeV LHC will be sensitive to the presence of the non-standard heavy $s$~scalar that is characteristic of the renormalizable coloron model.

\section*{Acknowledgments}

R.S.C. and E.H.S. are supported, in part, by the U.S. National Science Foundation under Grant No. PHY-0854889. During the completion of this work, A.F. was supported in part by the Tsinghua Outstanding Postdoctoral Fellowship, and by the NSF of China (under grants 11275101, 11135003). J.R. is supported by the China Scholarship Council, and by the NSF of China (under grants 11275101, 10625522, 11135003).

\end{document}